\documentclass[english,aps,prc,nofootinbib,superscriptaddress,twocolumn,letter]{revtex4}
\usepackage[latin9]{inputenc}
\usepackage{hyperref}
\hypersetup{colorlinks=true, linkcolor=blue, citecolor=blue, urlcolor=blue}

\usepackage{breakurl}
\usepackage{graphicx}
\usepackage{epsf}
\usepackage{epsfig}
\usepackage{amssymb,amsmath}
\usepackage[usenames]{color}
\usepackage{amssymb}
\usepackage{times}
\usepackage{comment}
\usepackage{txfonts}

\usepackage[normalem]{ulem} 
\usepackage{float} 

\usepackage{graphicx}
\usepackage{hyperref}

\allowdisplaybreaks



\begin{document}

\preprint{This line only printed with preprint option}

\title{Bayesian inference of the magnetic field and chemical potential on holographic jet quenching in heavy-ion collisions}

\author{Liqiang Zhu}
\email{zhuliqiang@mails.ccnu.edu.cn}
\affiliation{Key Laboratory of Quark and Lepton Physics (MOE) \& Institute of Particle Physics, Central China Normal University, Wuhan 430079, China}

\author{Zhan Gao}
\email{gaozhan2@mails.ccnu.edu.cn}
\affiliation{Key Laboratory of Quark and Lepton Physics (MOE) \& Institute of Particle Physics, Central China Normal University, Wuhan 430079, China}

\author{Weiyao Ke}
\email{weiyaoke@ccnu.edu.cn}
\affiliation{Key Laboratory of Quark and Lepton Physics (MOE) \& Institute of Particle Physics, Central China Normal University, Wuhan 430079, China}

\author{Hanzhong Zhang}
\email{zhanghz@mail.ccnu.edu.cn}
\affiliation{Key Laboratory of Quark and Lepton Physics (MOE) \& Institute of Particle Physics, Central China Normal University, Wuhan 430079, China}

\begin{abstract}

Jet quenching is studied in a background magnetic field and a finite baryon chemical potential. The production of energetic partons is calculated using the next-to-leading order (NLO) perturbative Quantum Chromodynamics (pQCD) parton model, while the parton energy loss formula is obtained from the AdS/CFT correspondence incorporating the magnetic field and baryon chemical potential effects. 
Using Bayesian inference, we systemically compare the theoretical calculations with experimental data for the nuclear modification factor $R_{AA}$ of the large transverse momentum hadrons in different centralities of nucleus-nucleus collisions at 0.2, 2.76 and 5.02 TeV, respectively. 
The form of the holographic calculation leads to a strong negative correlation between the magnetic field and the chemical potential for a fixed amount of energy loss. This degeneracy can also be observed after the model calibration. Finally, we discussed the sensitivity of jet quenching phenomena to the enteral magnetic field and a background baryon chemical potential. 


\end{abstract}
\maketitle
\date{\today}

\section{Introduction}
\label{emsection1}

Ultra-relativistic heavy-ion collisions conducted at the Relativistic Heavy Ion Collider (RHIC) and the Large Hadron Collider (LHC) are thought to have generated a novel state of matter referred to as the Quark Gluon Plasma (QGP)~\cite{PHENIX:2004vcz, STAR:2010vob}. Investigating particles with high momentum is an essential method for elucidating the characteristics of QGP. These high-energy particles are produced by high-energy partons, the creation of which can be accurately calculated from perturbative Quantum Chromodynamics (pQCD)~\cite{Gyulassy:2003mc, Baier:2000mf, Majumder:2010qh, dEnterria:2009xfs}, thereby rendering these hard probes highly effective for examining the properties of the medium.
An essential feature of the medium-parton interaction is the phenomenon of energy loss in high energy nucleus-nucleus collisions, resulting in a suppressed hadron cross-section at high transverse momentum ($p_T$) relative to what is anticipated by a direct scaling of the cross-section in proton-proton collisions. This effect, termed jet quenching~\cite{PHENIX:2001hpc, STAR:2003fka, BRAHMS:2003sns, Li:2024uzk, Xie:2024xbn}, is most effectively analyzed using the nuclear modification factor $R_{AA}$. $R_{AA}$ is defined as the ratio of particle spectrum in heavy-ion collisions to the product of proton-proton cross-section and the nucleus-nucleus thickness overlapping function. Its deviation from unity indicates nuclear medium effects.

A commonly applied approach to calculate energy loss in QGP relies on the premise of weakly coupled scenario between high-energy partons and the QGP. 
Within this scenario, high-energy partons traverse the medium along the light cone, primarily losing energy via medium-induced gluon emission as a consequence of multiple collisions with the medium. 
This weakly coupled approach has been remarkably effective in describing the suppression of $R_{AA}$ as observed in nuclear collisions. 
However, in many of these models, elastic energy loss are perturbatively calculated and receives large contributions from soft momentum $q\sim g_s T$ exchange with the medium, where $g_s$ is the QCD coupling.
This is hard to reconcile with the experimental discovery that the QGP produced in heavy-ion collisions is strongly coupled~\cite{Heinz:2013th, Ryu:2015vwa}. In fact, the phenomenological coupling is very large that $g_s T$ is almost comparable to the thermal kinetic energy. Therefore, this weak coupling approach may have certain limitations when dealing with strongly coupled QGP.
As a result, the question of whether jet quenching can be understood from a non-perturbative or strongly coupled perspective is an interesting one. In this paper, we adopt a strongly coupled approach, assuming strong coupling between the medium and parton, at the same time, introduce the dependence on background magnetic field and chemical potential, and investigate jet quenching phenomena using gauge/gravity duality~\cite{tHooft:1993dmi, Susskind:1994vu}.

Gauge/gravity dualities represent a broad collection of concepts asserting that gauge field theories in four-dimensional flat space are dual to gravity theories in curved space with an additional dimension. The Anti-de Sitter/Conformal Field Theory ($\mathrm{AdS/CFT}$) correspondence~\cite{Maldacena:1997re, Gubser:1998bc, Witten:1998qj, Aharony:1999ti} is a specific instance of gauge/gravity dualities. Since its introduction in the late 1990s, it has emerged as one of the most intensively researched areas in theoretical high-energy physics. 
$\mathrm{AdS/CFT}$ correspondence means that there is a certain relationship between the $\mathrm{AdS}_{5}\times S^{5}$ space-time type $\mathrm{IIB}$ string theory and the $\mathcal{N}$= 4 $\mathrm{SYM}$ (Supersymmetric Yang-Mills) gauge field theory in the (3+1)-dimensional Minkowski space-time. While $\mathcal{N}$= 4 $\mathrm{SYM}$ at zero temperature exhibits distinct differences from $\mathrm{QCD}$ in various aspects, it can still provide insights into certain qualitative characteristics of $\mathrm{QCD}$ under the strongly coupled regime at finite temperature. Within the AdS/CFT correspondence, the features of the medium are embedded in the background metric of the associated string theory (e.g., the $\mathrm{AdS}_{5}$ metric for the case of $\mathcal{N}$= 4 $\mathrm{SYM}$), whereas the traits of diverse dynamic processes occurring within the medium are manifested through the behavior of classical strings in this background. According to the strong and weak correspondences in $\mathrm{AdS/CFT}$, the strong coupling problem faced by four-dimensional conformal field theory can be solved by the corresponding weak coupling method of superstring theory. Up to now, researchers have exploited the $\mathrm{AdS/CFT}$ to explore various properties of $\mathrm{QGP}$. For example, jet quenching parameters~\cite{Liu:2006ug, Li:2016bbh, Zhang:2018pyr, Chen:2024epd}, the ratio of shear viscosity to entropy density~\cite{Critelli:2014kra}, phase transition~\cite{Cao:2024jgt, Zhu:2025gxo, Chen:2024mmd, Chen:2024ckb, Chen:2024jet, Bu:2024fhz, Zhao:2023gur, Zhao:2022uxc}, heavy quark potential~\cite{Rougemont:2014efa, Zhang:2018fpe, Luo:2024iwf, Guo:2024qiq}, heavy quark energy loss~\cite{Noronha:2010zc, Herzog:2006gh, Gubser:2006bz, Dai:2025dir} and light quark energy loss~\cite{Ficnar:2013qxa, Ficnar:2013wba, Ficnar:2014pmc, Zhu:2019ujc, Li:2025ugv, Zhang:2025wxi}.

This paper will focus on the effects of the constant magnetic field and chemical potential on holographic jet energy loss in high energy nucleus-nucleus collisions at RHIC/LHC. Recent researches hasve indicated that in the initial stages of noncentral ultra-relativistic heavy ion collisions at LHC energies, extremely high magnetic fields may be generated with magnitudes on the order of $eB\sim 70m_{\pi}^{2}$~\cite{Deng:2012pc, Huang:2015oca, Pang:2016yuh, Jiang:2022uoe}. The generated magnetic field decreases rapidly but is still influential in the initial stage of $\mathrm{QGP}$ formation 
~\cite{Deng:2012pc,Endrodi:2024cqn}.
This strong magnetic field affects the QCD phase transition~\cite{Bali:2011qj,Ding:2020inp,Ding:2024sux}, plasma evolution, and charge dynamics in strongly interacting matter~\cite{Astrakhantsev:2019zkr,Almirante:2024lqn,Ding:2025jfz}. 
The effect of a finite baryon chemical potential is small at LHC energy but can be influential at various RHIC energies~\cite{Andronic:2017pug}. 
Since both the magnetic and chemical potential might alter the properties of the partonic degrees of freedom of the QGP, it is interesting to study the corresponding response of jet quenching.
Recent work has separately studied the effects of magnetic fields and chemical potential on parton energy loss~\cite{Zhang:2019jfq, Hou:2021own}. In this paper, we simultaneously analyze the effects of both the magnetic field and chemical potential, revealing their strong correlations on impacting parton energy loss. Finally, using using Bayesian inference, we discuss the phenomenological sensitivity of jet quenching observables to magnetic field and chemical potential.

The paper is organized as follows: In Sec.\uppercase\expandafter{\romannumeral2}, we simply review the NLO pQCD parton model. A holographic model of the energy loss incorporating magnetic fields and chemical potentials is introduced in Sec.\uppercase\expandafter{\romannumeral3}. In Sec.\uppercase\expandafter{\romannumeral4}, we determine the coupling constant  in the model. In Sec.\uppercase\expandafter{\romannumeral5} we describe the holographic jet energy loss in the background of a magnetic field and chemical potential. Sec.\uppercase\expandafter{\romannumeral6} shows details of the Bayesian analysis, and results are discussed in Sec.\uppercase\expandafter{\romannumeral7}. Finally, we give a summary and an outlook in Sec.\uppercase\expandafter{\romannumeral8}.

\section{An NLO pQCD parton model}
\label{section4}

In this section, we will provide a brief overview of the $\mathrm{pQCD}$ computations concerning the next-to-leading-order $(\mathrm{NLO})$ cross sections for the production of single inclusive hadrons at high transverse momentum $p_{T}$ in $p+p$ and $A+A$ collisions, utilizing the concept of collinear factorization.

\subsection{Factorized calculations in $p + p$ collisions
\label{section4_sub1}
}

The invariant cross section for producing a single hadron with high transverse momentum in high-energy collisions can be factorized within the parton model. This factorization involves the convolution of collinear parton distribution functions (PDFs), hard scattering cross sections, and fragmentation functions (FFs)~\cite{Aversa:1988vb}. The differential cross section as a function of the hadron with transverse momentum $p_{T}$ and rapidity $y_{h}$ is~\cite{Xie:2022fak}:
\begin{align}
\begin{split}
\frac{d\sigma_{pp}^{h}}{dy_h d^{2}p_{T}}&=\sum_{abcd}\int dx_{a}dx_{b}f_{a/p}(x_{a},\mu^{2})f_{b/p}(x_{b},\mu^{2})\\
&\quad\times\frac{1}{\pi}\frac{d\sigma_{ab\rightarrow cd}}{d\hat{t}}\frac{D_{c}^{h}(z_{c},\mu^{2})}{z_{c}}+\mathcal{O}(\alpha_{s}^{3}) .
\end{split}
\label{eq.ppCrossSection}
\end{align}
Here, $f_{a/p}(x_{a},\mu^{2})$ and $f_{b/p}(x_{b},\mu^{2})$ are the parton distribution functions obtained from $\mathrm{CT18}$~\cite{Hou:2019efy} parametrization, while $D_{c}^{h}(z_{c},\mu^{2})$ is the fragmentation function of Kniehl-Kramer-Potter parametrization in vacuum from Ref~\cite{Kniehl:2000fe}. 
The LO partonic cross section for the $2\rightarrow2$ process $ab\rightarrow cd$ is denoted by $d\sigma_{ab\rightarrow cd}$.
The NLO correction at $\mathcal{O}(\alpha_{s}^{3})$ contains virtual corrections to the $2\rightarrow2$ cross sections and $2\rightarrow3$ tree-level cross sections. We take the renormalization scale of $\mu=1.2\;p_{T}^{h}$ at RHIC and LHC to describe hadron production in vacuum.

\begin{figure}[ht]
\begin{center}
\includegraphics[width=\columnwidth]{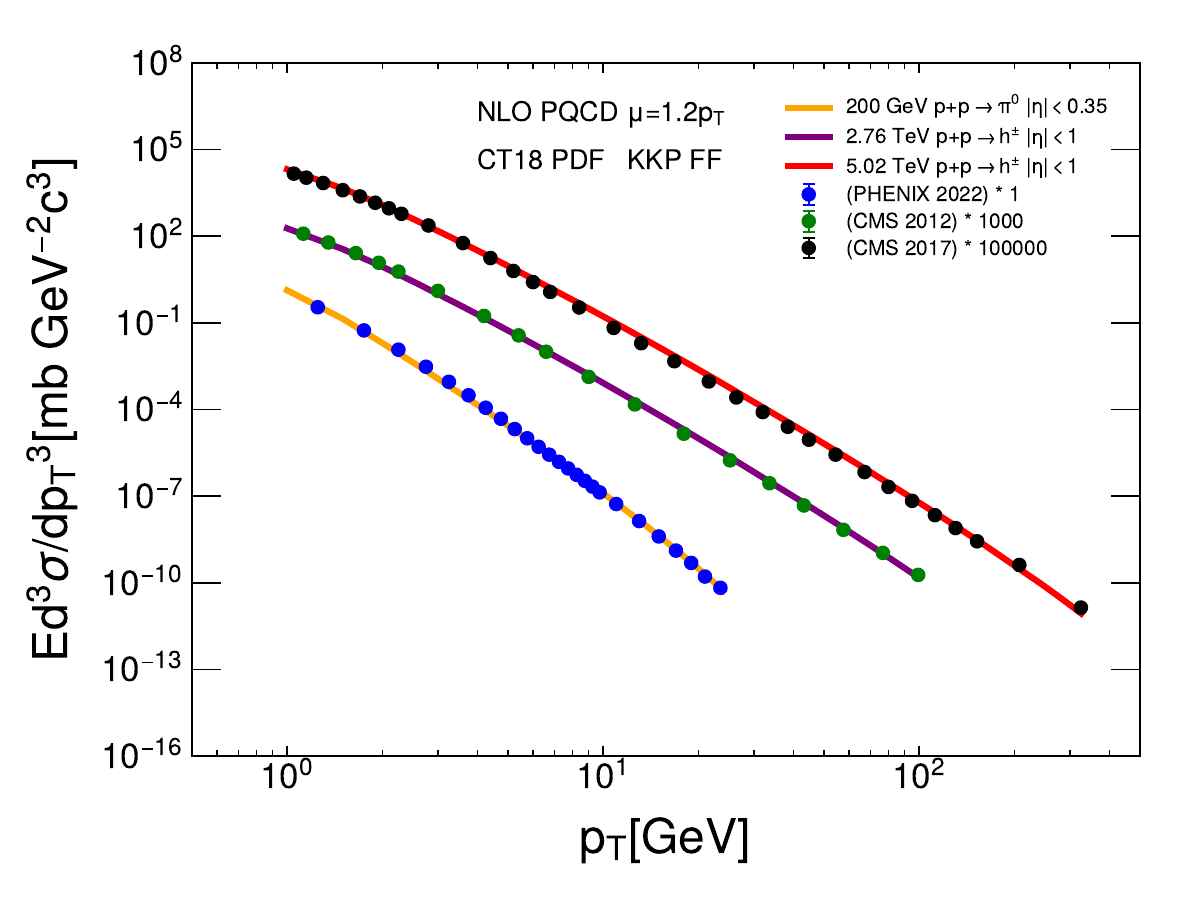} 
\end{center}
\caption{Cross sections of the single hadron production in $p+p$ collision compared with experiment data ~\cite{PHENIX:2021dod, CMS:2012aa, CMS:2016xef}.}
\label{fig.pphadron} 
\end{figure}

\autoref{fig.pphadron} shows the NLO pQCD result on the hadron production cross section in $p+p$ collisions compared with experimental data~\cite{PHENIX:2021dod, CMS:2012aa, CMS:2016xef}.
These numerical results show that the NLO pQCD parton model give a good description of the experimental data of single-hadron production at large $p_{T}$ in $p+p$ collision.

\subsection{Factorized calculations in $A + A$ collisions
\label{section4_sub2}
}

In nucleus-nucleus collisions with a fixed impact parameter $\Vec{b}$ , the single-inclusive hadron production spectra at high transverse momentum $p_{T}$ can be expressed as (analogous to Eq. \ref{eq.ppCrossSection})~\cite{Chen:2010te, Liu:2015vna, Zhang:2007ja}:
\begin{align}
\begin{split}
\frac{dN_{AB}^{h}(b)}{dyd^{2}p_{T}}&=\sum_{abcd}\int dx_{a}dx_{b}d^{2}rt_{A}(\vec{r})t_{B}(\vec{r}+\vec{b})\\
&\quad\times f_{a/A}(x_{a},\mu^{2},\vec{r})f_{b/B}(x_{b},\mu^{2},\vec{r}+\vec{b})\\
&\quad\times \frac{1}{\pi}\frac{d\sigma_{ab\rightarrow cd}}{d\hat{t}}\frac{D_{c}^{h}(z_{c},\mu^{2},\bigtriangleup E_{c})}{z_{c}}+\mathcal{O}(\alpha_{s}^{3}) .
\end{split}
\end{align}
Here, $t_{A}(\vec{r})$ and $t_{B}(\vec{r}+\vec{b})$  are the projectile and target nuclear thickness functions,
normalized as $\int d^{2}rt_{A}(\vec{r})=A$ with A the mass number of the nucleus. Here we use the Woods-Saxon form for the nuclear density distribution. $f_{a/A}(x_{a},\mu^{2},\vec{r})$ is the nuclear modified PDF~\cite{Wang:1996yf, Wang:1998ww, Li:2001xa, Chen:2008vha}:

\begin{align}
\begin{split}
f_{a/A}(x_{a},\mu^{2},\vec{r})&=S_{a/A}(x_{a},\mu^{2},\vec{r})\bigg[\frac{Z}{A}f_{a/p}(x_{a},\mu^{2})\\
&\quad+\bigg(1-\frac{Z}{A}f_{a/n}(x_{a},\mu^{2}\bigg)\bigg] ,
\end{split}
\end{align}
where $Z$ is the proton number of the nucleus. $S_{a/A}(x_{a},\mu^{2},\vec{r})$ is called the nuclear shadowing factor and denotes the modification of nuclear PDF as compared to the simple isospin average using the PDF of a free proton $f_{a/p}(x_{a},\mu^{2})$. Here, the shadowing factor $S_{a/A}(x_{a},\mu^{2},\vec{r})$ takes the following form~\cite{Zhou:2010zzm, Emelyanov:1999pkc, Hirano:2003pw}
\begin{align}
S_{a/A}(x_{a},\mu^{2},\vec{r})&=1+A\frac{t_{A}(\vec{r})\big[S_{a/A}(x_{a},\mu^{2})-1\big]}{\int d^{2}r[t_{A}(\vec{r})]} ,   
\end{align}
and we use the $\mathrm{EPPS21}$ parametrization for $S_{a/A}(x_{a},\mu^{2})$~\cite{Eskola:2021nhw}.
Finally, $D_{h/c}(z_{c},\mu^{2},\Delta E_{c})$ is the medium-modified fragmentation function, given by~\cite{Salgado:2003gb, Zhang:2007ja, Zhang:2009rn, Wang:2004yv, Gyulassy:2001nm}:

\begin{align}
D_{h/c}(z_{c},\mu^{2},\Delta E_{c})&=\frac{z_{c}^{\prime}}{z_{c}}D_{h/c}(z_{c}^{\prime},\mu^{2}) ,
\end{align}
where $\Delta E_{c}$ is the energy loss of parton $c$. The variable $z_{c}=p_{T}/p_{Tc}$ represents the vacuum fragmentation momentum fraction of a hadron from parton $c$, while $z_{c}^{\prime}=p_{T}/(p_{Tc}-\Delta E_{c})$ corresponds to the in-medium case, where the parton loses energy $\Delta E_{c}$ prior to fragmentation. The calculation of $\Delta E_{c}$ will be given in the next section from the holographic model.

\subsection{Nuclear modification factors
\label{section4_sub3}
}

Based on the calculations in both $p+p$ and $A+A$ collisions as discussed above, the nuclear modification factor for single-inclusive hadron production in heavy-ion collisions can be computed following the approach in~\cite{Muller:2012zq}:

\begin{align}
R_{AA}(p_{T})=\frac{\frac{dN_{AA}}{dyd^{2}p_{T}}}{T_{AB}(\vec{b})\frac{d\sigma_{pp}}{dyd^{2}p_{T}}} ,
\end{align}

Here, $T_{AA}(\vec{b})=\int d^{2}\vec{r}t_{A}(\vec{r})t_{A}(\vec{r}+\vec{b})$ defines the nuclear overlap function, which quantifies the geometric overlap of two colliding nuclei at a specific impact parameter $\vec{b}$ for the specific centrality.


\section{Holographic model of the energy loss}
\label{section2}

\subsection{General set up}

\begin{figure}
\centering
\includegraphics[width=0.35\textwidth]{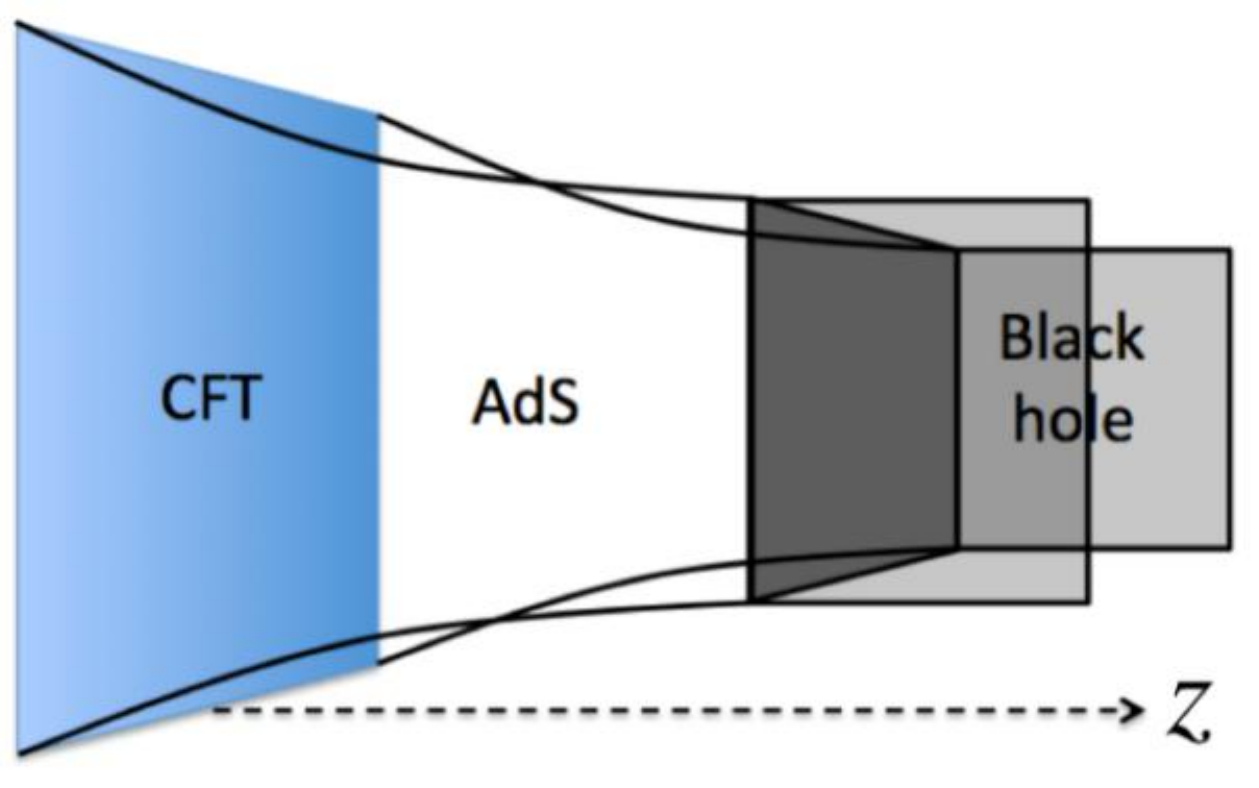}
\caption{Illustration of the $\mathrm{AdS/CFT}$ correspondence. $z$ is the coordinate into the bulk.}
\label{fig:AdS}
\end{figure}

Now we introduce a holographic model with a magnetic field and chemical potential. Within the $\mathrm{AdS/CFT}$ correspondence, introducing magnetic field and chemical potential into $\mathcal{N}$= 4 $\mathrm{SYM}$ can be achieved by endowing the black hole in the holographic dimension with
charge. The resulting spacetime geometry is described by an AdS-RN black hole, whose dynamics are governed by the following action\cite{Chamblin:1999tk}
\begin{align}\label{eq7}
I=\frac{1}{2\kappa^{2}}\int d^{5}x\sqrt{-g}(\mathcal{R}+\frac{12}{L^{2}}-\frac{L^{2}}{g^{2}_{F}}F_{\mu\nu}F^{\mu\nu}) ,
\end{align}
In this context, $\kappa_{4}^{2}=8\pi G$, where $G$ is the gravitational constant, and $\mathcal{R}$ represents the Ricci scalar. The parameter $L$ denotes the radius of the $\mathrm{AdS}$ space, which, for simplicity, is normalized to unity ($L=1$) in the subsequent analysis. The effective dimensionless gauge coupling constant is denoted by $g_{F}$. The field strength tensor $F_{\mu\nu}$ is expressed as $F_{\mu\nu}=\partial_{\mu}A_{\nu}-\partial_{\nu}A_{\mu}$, with $A_{\mu }$ being the $U(1)$ gauge field. The 5-dimensional solution to the equations of motion derived from Eq.~(\ref{eq7}) is given by:
\begin{align}\label{eq8}
ds^{2}=\frac{1}{z^{2}}\bigg(-f(z)dt^{2}+d\vec{x}^{2}+\frac{dz^{2}}{f(z)}\bigg) ,
\end{align}
with
\begin{align}\label{eq9}
\begin{split}
f(z)=1-(1+Q^{2})\bigg(\frac{z}{z_{h}}\bigg)^{4}+Q^{2}\bigg(\frac{z}{z_{h}}\bigg)^{6} ,
\end{split}
\end{align}
where $Q^{2}=\mu^{2}_{B}z_{h}^{2}+B^{2}z_{h}^{4}$ \cite{Hartnoll:2007ai, Hartnoll:2009sz, Rodrigues:2020ndy}, which is the charge of the black hole, $\mu_B$ and $B$ are the baryon chemical potential and background magnetic field. $t$ is the time coordinate and $\vec{x}$ is the $\mathrm{CFT}$ space coordinates on the boundary. $z$ is the $\mathrm{AdS}$ space coordinate, and $z=z_h$ is the horizon which near the boundary of  black hole 
, as shown in figure~\ref{fig:AdS}.

We use the Hawking formula of the black hole
\begin{align}\label{eq10}
T(z_{h},\mu_{B},B)=\frac{1}{\pi z_{h}}(1-\frac{Q^{2}}{2}).
\end{align}
It should be noted that for a given set of $T$, $\mu_{B}$ and $B$, Eq.~(\ref{eq10}) yields four roots for $z_{h}$. 
However, the physically acceptable solution is the only one of these roots that is real and positive.
Therefore, in the following, we will only consider the branch with $z_{h}>0$. 

\subsection{Energy loss in magnetic field and chemical potential background}
\label{section3}

In this section, we apply the methodology presented in~\cite{Ficnar:2013qxa, Ficnar:2013wba} to investigate the impact of a magnetic field and chemical potential on the energy loss of light quarks using finite endpoint momentum shooting strings. In this approach, a specific classical string motion is considered, where the endpoint of the string starts near the horizon and moves towards the boundary, while carrying certain energy and momentum. As the string rises, this energy and momentum gradually dissipate into the remaining part. Hence, this type of motion is called a finite-endpoint-momentum shooting string. 


The AdS space-time metric Eq.~(\ref{eq8}) can be rewritten in the form (here, $dx$ represents $d\vec{x}$ in Eq.~(\ref{eq8})):
\begin{align}\label{eq11}
ds^{2}&=G_{tt}(z)dt^{2}+G_{xx}(z)dx^{2}+G_{zz}(z)dz^{2},
\end{align}
where
\begin{align}\label{eq12}
G_{tt}(z)&=-\frac{1}{z^{2}}f(z), \; G_{xx}(z)=\frac{1}{z^{2}}, \; G_{zz}(z)=-\frac{1}{z^{2}}\frac{1}{f(z)},
\end{align}
The following derivation can be directly extended to a wider class of metrics beyond the present case; however, Eq.~(\ref{eq11}) already captures many interested scenarios. 
Because the metric does not exhibit explicit dependence on $t$ or $x$, the quantity defined below remains conserved along geodesic paths (adopting the notation convention from~\cite{Ficnar:2013wba}):
\begin{align}\label{eq13}
R=\frac{G_{tt}(z)dt}{G_{xx}(z)dx},
\end{align}
then,
\begin{align}\label{eq14}
dt^{2}=\bigg(\frac{RG_{xx}(z)dx}{G_{tt}(z)}\bigg)^{2}.
\end{align}
The finite momentum endpoints will follow $R$-parametrized null geodesics with $ds^{2}=0$ :
\begin{align}\label{eq15}
G_{tt}(z)dt^{2}+G_{xx}(z)dx^{2}+G_{zz}(z)dz^{2}=0 .
\end{align}
Substituting Eq.~(\ref{eq14}) into Eq.~(\ref{eq15}), we have
\begin{align}\label{eq16}
dx^{2}\bigg(\frac{R^{2}G_{xx}(z)}{G_{tt}(z)}+G_{xx}(z)\bigg)+G_{zz}(z)dz^{2}=0,
\end{align}
which solves to 
\begin{align}\label{eq17}
\nonumber
\bigg(\frac{dx}{dz}\bigg)^{2}&=-\frac{G_{tt}(z)G_{zz}(z)}{G_{xx}(z)\big[G_{tt}(z)+G_{xx}(z)R^{2}\big]}\\
\nonumber
& =-\frac{-\frac{L^{2}}{z^{2}}f(z)\frac{L^{2}}{z^{2}}\frac{1}{f(z)}}{\frac{L^{2}}{z^{2}}\big(-\frac{L^{2}}{z^{2}}f(z)+\frac{L^{2}}{z^{2}}R^{2}\big)}\\
& =\frac{1}{R^{2}-f(z)}
\end{align}
The geodesic cannot extend beyond a (minimum) $z=z_{*}$ at which the denominator of Eq.~(\ref{eq17}) vanishes, where $z_{*}$is a very small value in the coordinates of the $\mathrm{AdS}$ space. It can be connected to $R$ if the geometry Eq.~(\ref{eq11}) permits null geodesics such that
\begin{align}\label{eq18}
G_{tt}(z_{*})=-G_{xx}(z_{*})R^{2},
\end{align}
obtianing, 
\begin{align}\label{eq19}
R^{2}=-\frac{G_{tt}(z_{*})}{G_{xx}(z_{*})}=f(z_{*}).
\end{align}
According to Eq.~(\ref{eq17}), we obtain
\begin{align}\label{eq20}
\frac{dx}{dz}=\frac{1}{\sqrt{R^{2}-f(z)}}=\frac{1}{\sqrt{f(z_{*})-f(z)}}
\end{align}
Integrating from $z$ to $z_h$, we get the relation between $x$ and $z$:
\begin{align}\label{eq21}
x=&\int_{z}^{z_{h}} \frac{dz}{\sqrt{\frac{z^3}{z_{h}^3}(1+\mu^{2}_{B}z_{h}^{2}+B^{2}z_{h}^{4})-\frac{z^4}{z_{h}^4}(\mu^{2}_{B}z_{h}^{2}+B^{2}z_{h}^{4})}} .     
\end{align}

Because components of the metric in Eq.~(\ref{eq11}) does not explicitly dependent on $t$, a straightforward formula in Ref.~\cite{Ficnar:2013wba} can be used to determine the flow of energy from the terminus to the majority of the string, 
\begin{align}\label{eq22}
\dot{p}_{t}=-\frac{1}{2\pi\alpha^{\prime}}G_{tt}(z)\dot{t},
\end{align}
where $\alpha^{\prime}$ is related to the string tension.
This equation means that
the energy drain from a finite-momentum endpoint is caused by string world-sheet currents, which know nothing about its finite momentum other than its existence via altered boundary conditions~\cite{Ficnar:2013wba}. Substituting Eq.~(\ref{eq13}) into Eq.~(\ref{eq22}) we get
\begin{align}\label{eq23}
\frac{dE}{dx}=\frac{\vert R\vert}{2\pi\alpha^{\prime}}G_{xx}(z),
\end{align}
It is important to mention that in the small $z_{*}$
limit for asymptotically AdS geometries, it is common to take $z_{*}=0$ ($R=1)$~\cite{Ficnar:2013qxa, Ficnar:2013wba}, where $z=z_{*}=0$ is the boundary of $\mathrm{AdS}$ space, and is further simplified to get:
\begin{align}
\frac{dE}{dx}=-\frac{\sqrt{\lambda}}{2\pi}\frac{1}{z^{2}} .
\label{eq24}
\end{align}
where $\sqrt{\lambda}=L^{2}/\alpha^{\prime}$ and $\lambda$ is the 't Hooft coupling constant.

Although the methods for calculating energy loss in \cite{Ficnar:2013qxa, Ficnar:2013wba} have been improved by subsequent research \cite{Chesler:2014jva, Chesler:2015nqz}, in this work, we are the first to incorporate chemical potential and magnetic field into the energy loss based on the methods described in \cite{Ficnar:2013qxa, Ficnar:2013wba}.

\subsection{Basic properties of the energy loss rate}

\begin{figure}
\centering
\includegraphics[width=0.9\columnwidth]{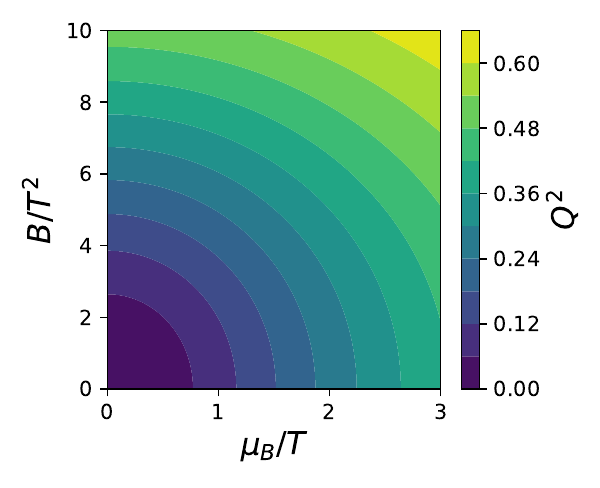}
\caption{The squared effective charge $Q^2$ as a function of the scaled baryon chemical potential and scaled magnetic field.}
\label{fig:scaled_zh_Q2}
\end{figure}

\begin{figure}
\centering
\includegraphics[width=0.9\columnwidth]{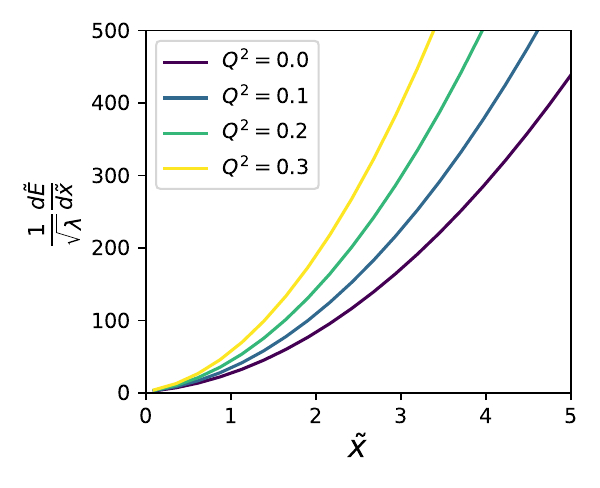}
\caption{The rescaled energy loss as functions of the rescaled path length $\tilde{x}$ and the external scale parameter $Q^2$.}
\label{fig:scaled_eloss}
\end{figure}

Before apply this formula to phenomenology, we first make general discussion on the $T$, $\mu_B$, and $B$ dependence of the energy loss per unit path length. Note that we can rescale all the variables using temperature as the natural units and define
\begin{align}
\nonumber
 \tilde{x} &= xT, \, \tilde{z} = zT, \, \tilde{z}_h = z_h T, \\
\tilde{E} &= E/T,\, \tilde{\mu}_B = \mu_B/T, \, \tilde{B} = B/T^2 .
\end{align}
In terms of the rescaled variables, we can rewrite the equation as 
\begin{align}
Q^2 &= \tilde{\mu}_B^2 \tilde{z}_h^2 + \tilde{B}^2 \tilde{z}_h^4, \,\,\,
\tilde{z}_h = \frac{2-Q^2}{2\pi}, \\
\tilde{x} & = \frac{\tilde{z}_h}{1+Q^2}\left(\sqrt{\left(1+Q^2\right)\frac{\tilde{z}_h^2}{\tilde{z}^2}-Q^2}-1\right), \\
\frac{1}{\sqrt{\lambda}}\frac{d\tilde{E}}{d\tilde{x}} &= -\frac{1}{2\pi} \frac{1}{\tilde{z}^2(\tilde{x}, Q^2)}.
\end{align}
With the first two equations, one can solve for the physical solution mentioned eariler $\tilde{z}_h = \tilde{z}_h(\tilde{\mu}_B, \tilde{B}) > 0$.
Then, we remark that all the information on the external energy scales ($B$ and $\mu_B$) only enters the energy loss rate via a unique combination $Q^2 = \tilde{\mu}_B^2 \tilde{z}_h^2(\tilde{\mu}_B, \tilde{B}) + \tilde{B}^2 \tilde{z}_h^4(\tilde{\mu}_B, \tilde{B})$. 
Figure~\ref{fig:scaled_zh_Q2} shows the iso-$Q^2$ lines as functions of $\tilde{\mu}_B$ and $\tilde{B}$. For a given $Q^2$, $\tilde{\mu}_B$ and $\tilde{B}$ are anti-correlated.
This means that there are degeneracy in the parameter space, i.e., knowing the energy loss rate alone cannot uniquely determine a set of $(\tilde{\mu}_B, \tilde{B})$. This observation will be reflected in our final results in Sec.~\ref{sec:Result}.
In figure~\ref{fig:scaled_eloss}, we plotted the scaled energy loss rate as functions of $Q^2$ and the scaled path length in the upper and lower panels respectively. The energy loss rate increases slightly with $Q^2$.

If all the external scale vanish $Q^2=0$, then the energy loss rate goes back to the formula used in Refs.~\cite{Ficnar:2013qxa, Ficnar:2013wba}
\begin{align}
\label{eq:eloss-0-0}
\left.\frac{1}{\sqrt{\lambda}}\frac{d\tilde{E}}{d\tilde{x}}\right|_{\tilde{\mu}_B=0, \tilde{B}=0} &= -\frac{\pi}{2} \left(\frac{1+\tilde{x}}{\tilde{z}_h(0,0)}\right)^2.
\end{align}
with $\tilde{z}_h(0,0)=1/\pi$.

\begin{figure}
\centering
\includegraphics[width=\columnwidth]{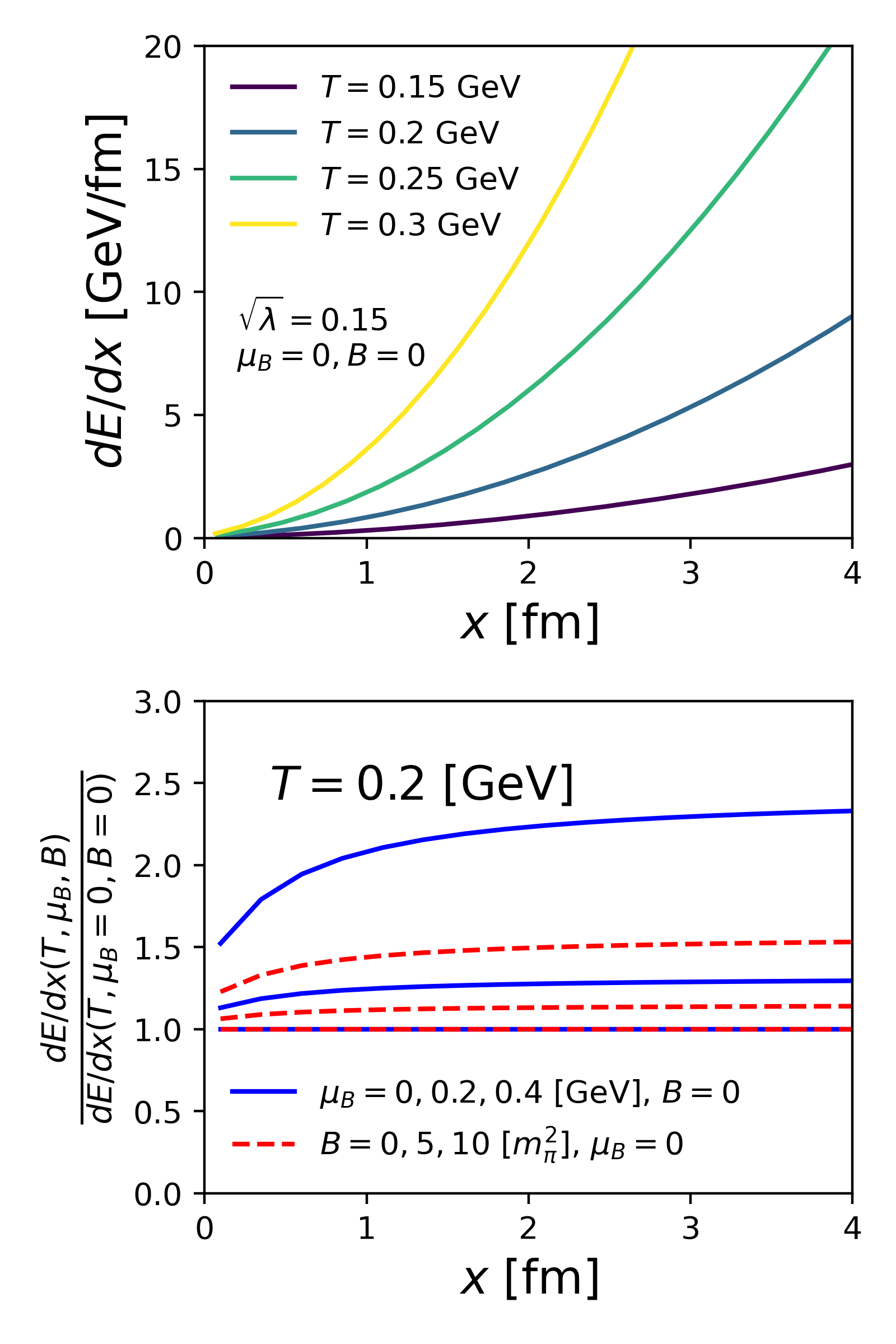}
\caption{Upper panel: The physical energy loss per unit path length at different temperatures with the magnetic field and baryon chemical potential set to zero and $\sqrt{\lambda}=0.15$. Lower panel: The effects of magnetic field and baryon chemical potential on energy loss at a fixed temperature of $T = 0.2$ GeV.}
\label{fig:physical_eloss}
\end{figure}

In figure~\ref{fig:physical_eloss}, we put in energy scales using different temperature and perform conversion of units. Here, we measure energy loss rates in GeV/fm, path length in fm, temperature and chemical potential in GeV, and magnetic fields in $m_\pi^2$ (setting the unit charge to be one).
In the upper panel of figure~\ref{fig:physical_eloss}, we show the relationship between the instantaneous energy loss rate $dE/dx$ and the distance $x$ at different temperatures under zero magnetic field and zero chemical potential. The results indicate that $dE/dx$ increases with temperature and path length. In fact, from Eq.~(\ref{eq:eloss-0-0}), one can deduces the limiting behavior at large $\tilde{x}=xT$ is $dE/dx\propto x^2 T^4$, which is very different from those given by perturbative calculations of collisional energy loss $dE_{\rm coll}/dx \propto g_s^4 T^2$.
In the lower panel of figure~\ref{fig:physical_eloss}, we investigated the effects of magnetic field and chemical potential on energy loss at a fixed temperature ($T=0.2\;\mathrm{GeV}$). The calculations demonstrate that both the magnetic field and chemical potential enhance $dE/dx$, suggesting they may be relevant for phenomenology at lower beam energies.

\section{Determine the phenomenological coupling $\lambda(T)$}
\begin{figure}
\begin{center}
\includegraphics[width=0.45\textwidth]{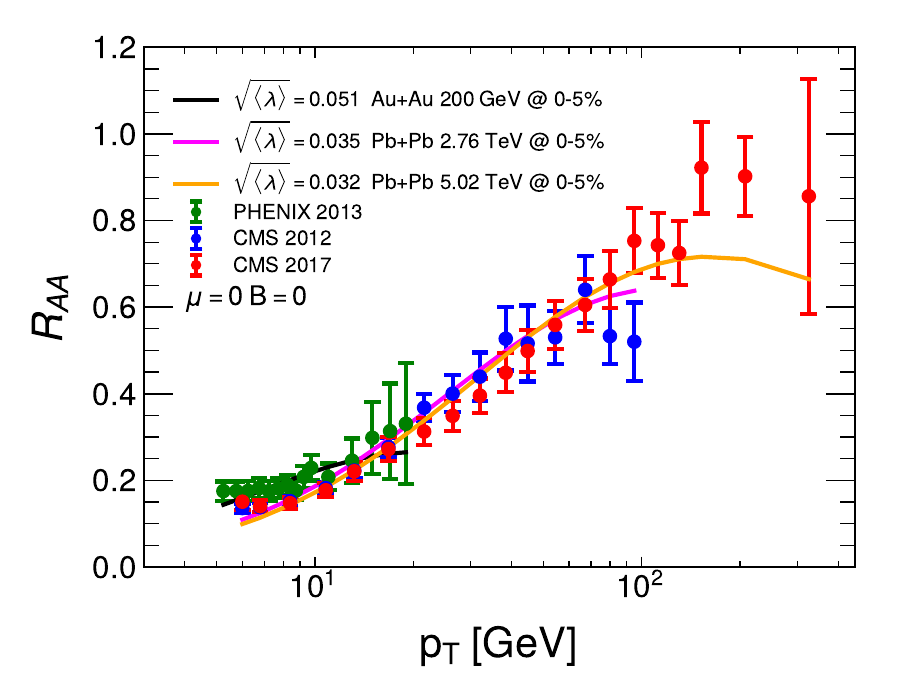}
\end{center}
\caption{The nuclear modification factor $R_{AA}$ for the best-fitting energy loss parameter $\left\langle\lambda\right\rangle$ in central A+A collisions at 200 GeV, 2.76 or 5.02 TeV, respectively. The experimental data are from~\cite{PHENIX:2012jha, CMS:2012aa, CMS:2016xef}.}
\label{fig:RAA0-5}
\end{figure}

We first set the magnetic field and chemical potential to zero, considering only the influence of $\lambda$ on energy loss. 
The temperature $T$ profile is obtained from the ClVisc simulation of viscous relativistic hydrodynamics~\cite{Pang:2018zzo,Wu:2018cpc}.
We assume that $\lambda$ is unchanged in a given center-of-mass energy. Under this assumption, $\lambda$ can be considered as an average value, recorded as $\left\langle\lambda\right\rangle$. We then calculated the nuclear modification factor $R_{AA}$ for central collisions in Au+Au at 200 GeV and Pb+Pb at 2.76 TeV and 5.02 TeV, and we use $\chi^{2}/d.o.f$ fitting of the calculated results against the experimental data to extract the optimal value of $\left\langle\lambda\right\rangle$. 
The definition of $\chi^{2}/d.o.f$ is as follows:
\begin{align}
\chi^{2}/d.o.f=\sum^{N}_{i=1}\bigg[ \frac{(V^{th}-V^{exp})^{2}}{\sum_{t}\sigma_{t}^{2}}\bigg]_{i}\bigg/N
\end{align}
Here, $V^{th}$ represents the results obtained from theoretical calculations,$V^{exp}$ represents the experimental results, $\sum_{t}\sigma_{t}^{2}$ denotes the sum of the squares of the different errors in the experimental data. N is the degree of freedom, representing the number of experimental data points. A smaller value of $\chi^{2}/d.o.f$ indicates better agreement between the theoretical calculations and the experimental data, while a larger value indicates worse agreement, respectively.
The results show that as the collision energy increases, and thus the temperature of QGP rises, $\left\langle\lambda\right\rangle$ decreases accordingly, as figure~\ref{fig:RAA0-5} shows.
In our calculations, the value of $\left\langle\lambda\right\rangle$  is shown to be small, indicating a weak coupling regime. However, we still employ the AdS/CFT correspondence in phenomenology.

\begin{figure}
\begin{center}
\includegraphics[width=\columnwidth]{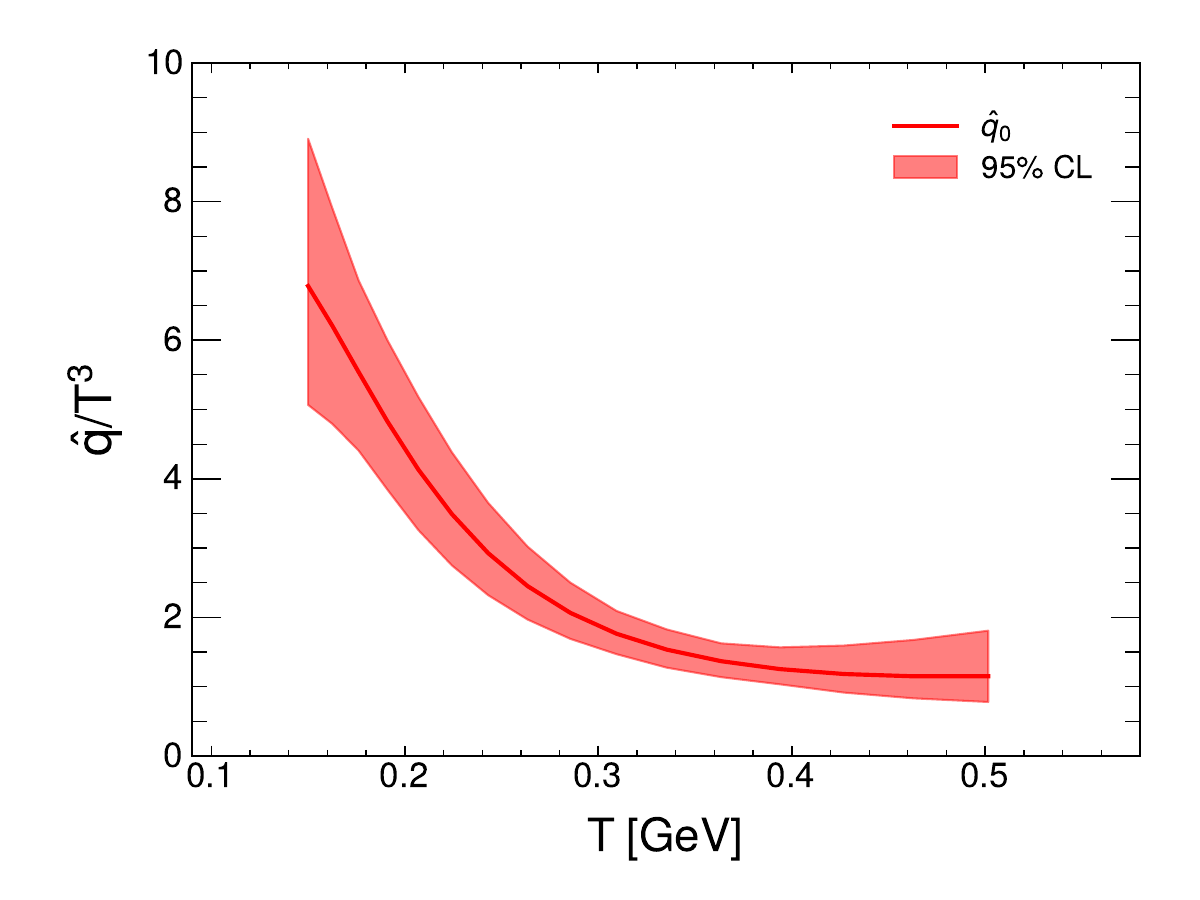} 
\includegraphics[width=\columnwidth]{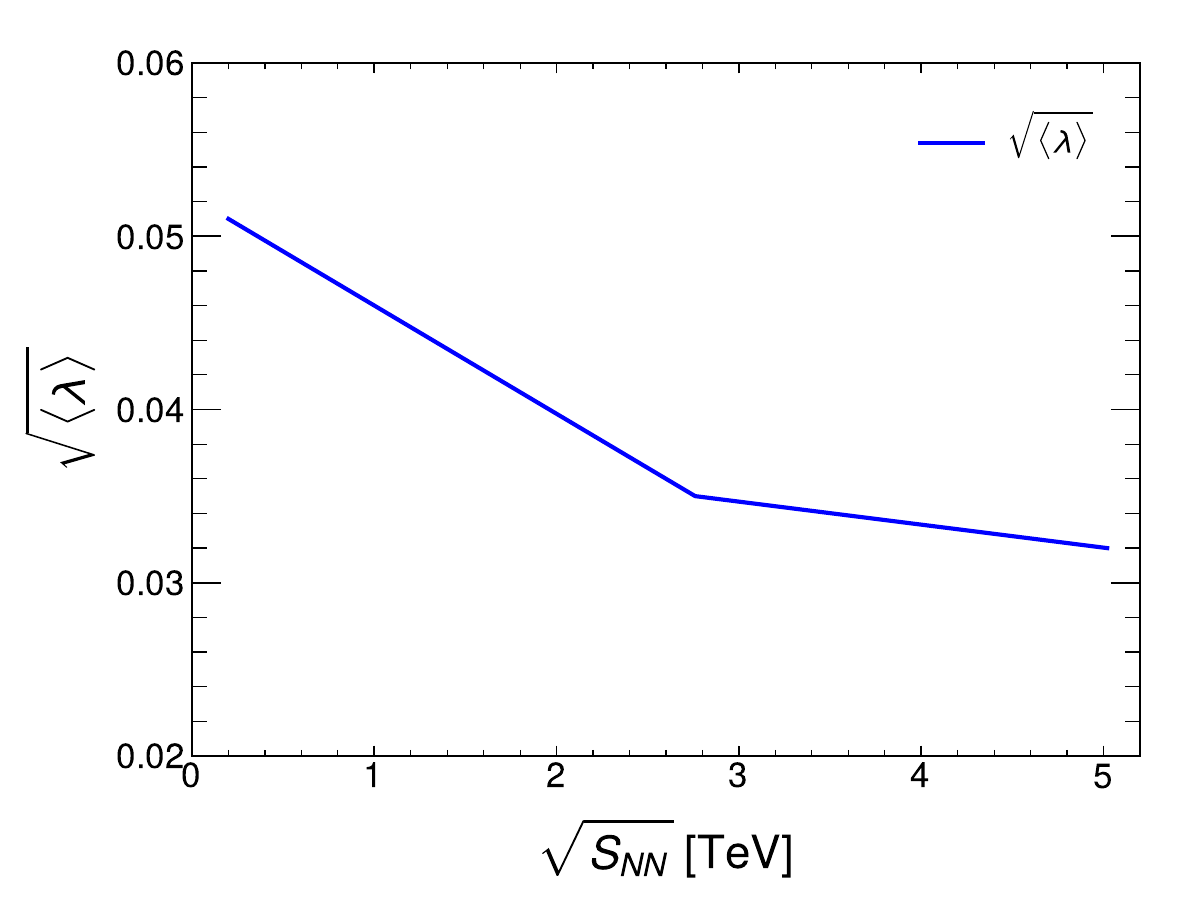} 
\end{center}
\caption{Upper panel: The dependence of $\hat{q}/T^{3}$ on temperature\cite{Xie:2022ght}. Lower panel: The dependence of $\sqrt{\langle\lambda\rangle}$ on the center-of-mass energy. 
}
\label{fig:lambdaAverage}
\end{figure}

In reference~\cite{Xie:2022ght}, a result showing that $\hat{q}/T^{3}$ decreases with the local temperature $T$ for a jet propagating through the QGP medium, as shown in the upper panel of figure~\ref{fig:lambdaAverage}. This result is consistent with the trend of the mass-center energy dependence of $\left\langle\lambda\right\rangle$ that we calculated above in figure~\ref{fig:RAA0-5},
as shown in the lower panel of figure~\ref{fig:lambdaAverage}, where $\sqrt{\left\langle\lambda\right\rangle}$ decreases with increasing center-of-mass energy.
Considering the medium temperature increases with the collision energy, we assume that the temperature dependence of $\sqrt{\left\langle\lambda\right\rangle}$ is proportional to the temperature dependence of $\hat{q}/T^{3}$, therefore, in the later part of the paper, we will adopt the temperature-dependent 't Hooft coupling $\lambda(T)$.

\begin{figure}
\begin{center}
\includegraphics[width=\columnwidth]{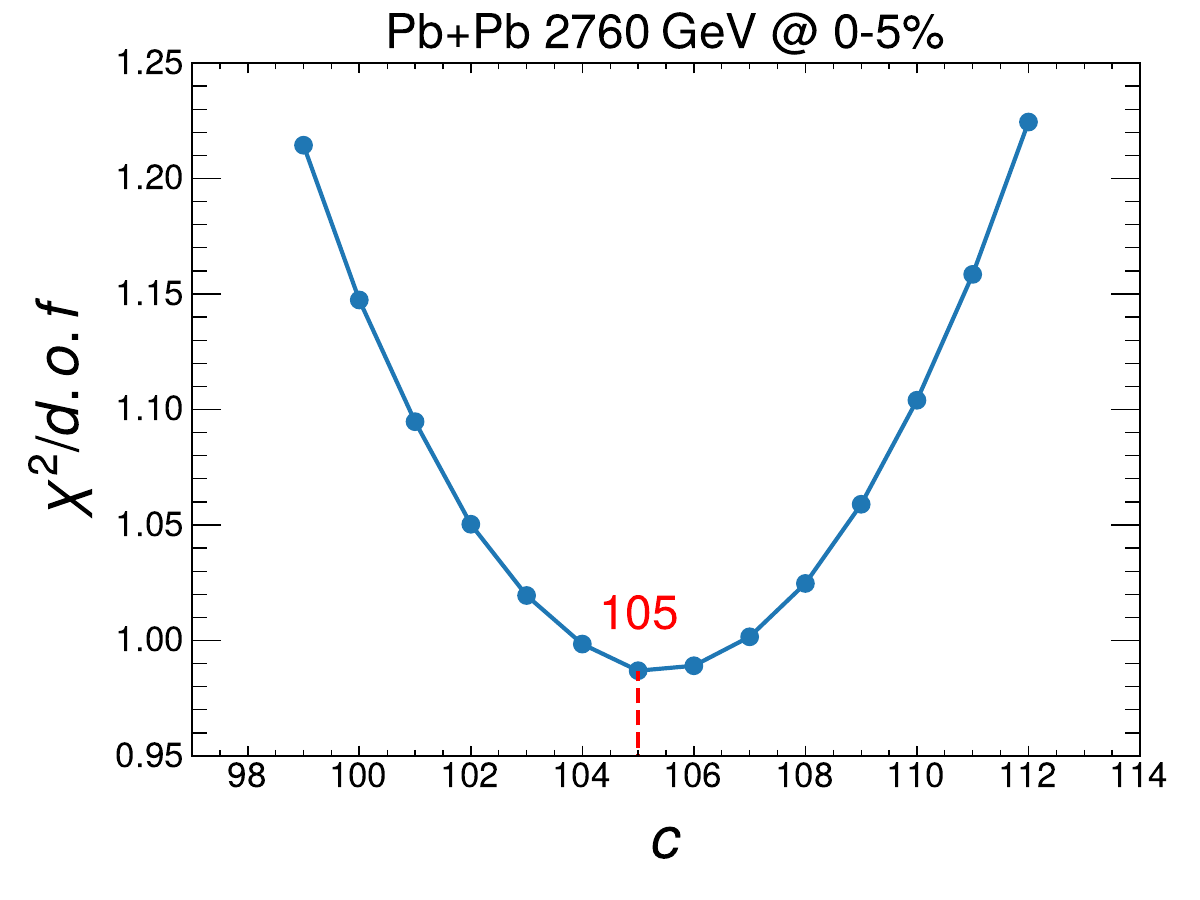} 
\includegraphics[width=\columnwidth]{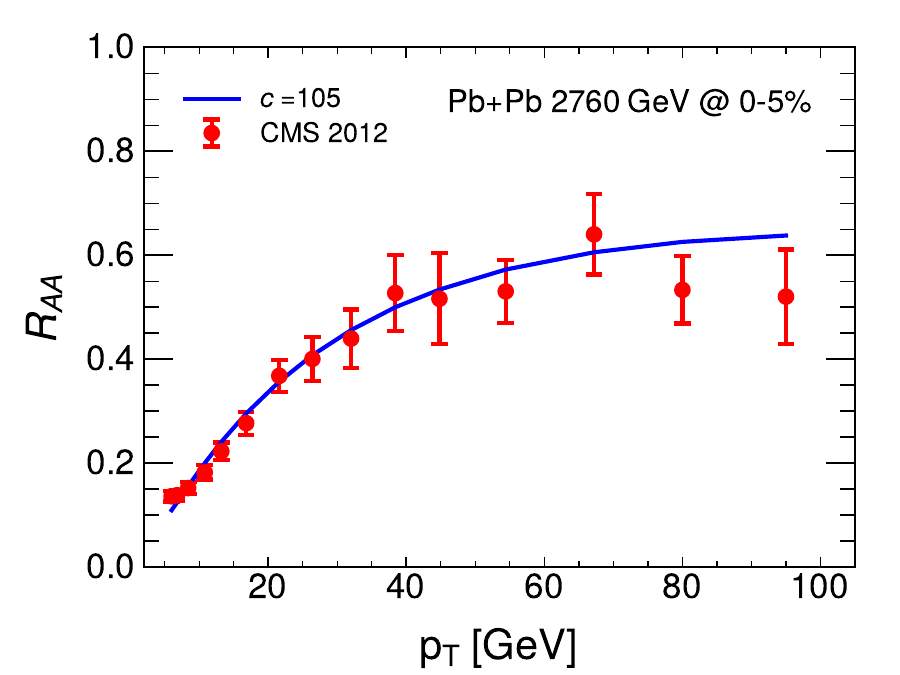} 
\end{center}
\caption{Upper panel: the $\chi^{2}/d.o.f$ fitting for the proportionality coefficient $c$ with the nuclear modification in 0-5\% Pb+Pb collisions at
$\sqrt{S_{NN}}=2.76$ TeV.
Lower panel: A comparison between the nuclear modification factor $R_{AA}$ calculated with the best-fitting value $c=105$ and the data~\cite{CMS:2012aa}.}
\label{fig.chi2ofc}
\end{figure}

In fact, in a hot $\mathcal{N}=4$ SYM theory a relationship between $\hat{q}/T^{3}$ and $\sqrt{\lambda}$ is obtained~\cite{Liu:2006ug}:
\begin{align}
\hat{q}_{\mathrm{SYM}}\sim\sqrt{\lambda}T^{3}.
\end{align}
Therefore, in the following study, we will parametrize the temperature dependence of the 't Hooft coupling by relating it to the temperature dependence of $\hat{q}/T^{3}$,
\begin{align}\label{eqn:lam-c}
\lambda(T)=\big(\hat{q}/T^{3}\big)\big/c,
\end{align}
where $\hat{q}/T^{3}$ as a function of $T$ is given by the reference~\cite{Xie:2022ght}, and $c$ is a tunable constant to be fixed by experiments.
Therefore, Eq. ($\ref{eqn:lam-c}$) is the temperature-dependent $\lambda(T)$.
To determine $c$, we consider that in central nucleus-nucleus collisions at the LHC energies (such as central Pb+Pb collisions at 2.76 or 5.02 TeV), the effect of magnetic field and chemical potential are both negligible, i.e., $B \to 0$ and $\mu_{B} \to 0$~\cite{Andronic:2017pug}. With jet energy loss by submitting Eq. ($\ref{eqn:lam-c}$) in Eq. ($\ref{eq24}$), we calculate the nuclear modification factor $R_{AA}$ for the central Pb+Pb collisions at 2.76 TeV (according to reference~\cite{Xie:2022ght}, the calculation results for Pb+Pb 2.76 TeV match the experimental data best, we utilized the result from the $\hat{q}_{0}$ line in the upper panel of figure~\ref{fig:lambdaAverage}). We then performed a $\chi^{2}/d.o.f$ fitting of the calculated results against the experimental data to extract the best-fitting constant $c$. 
Figure~\ref{fig.chi2ofc} shows the values of $\chi^{2}/d.o.f$ corresponding to different values of the constant $c$. As shown in figure~\ref{fig.chi2ofc}, when $c$=105, the agreement with the experimental data is the best. 
Therefore, in the following calculations, the value of the constant $c$ in Eq. ($\ref{eqn:lam-c}$) is fixed at 105.


\begin{figure*}
\begin{center}
\includegraphics[width=0.32\textwidth]{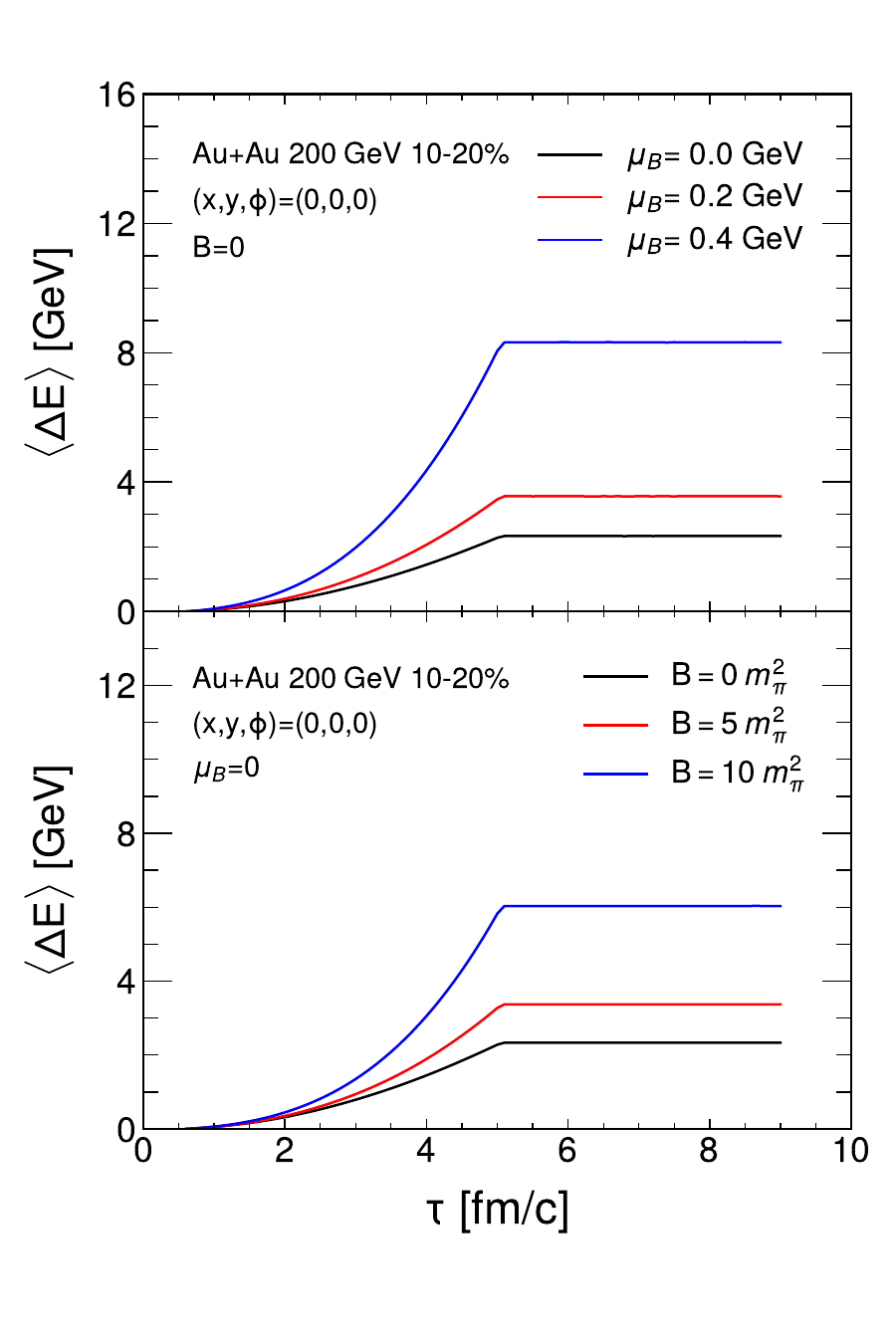}
\includegraphics[width=0.32\textwidth]{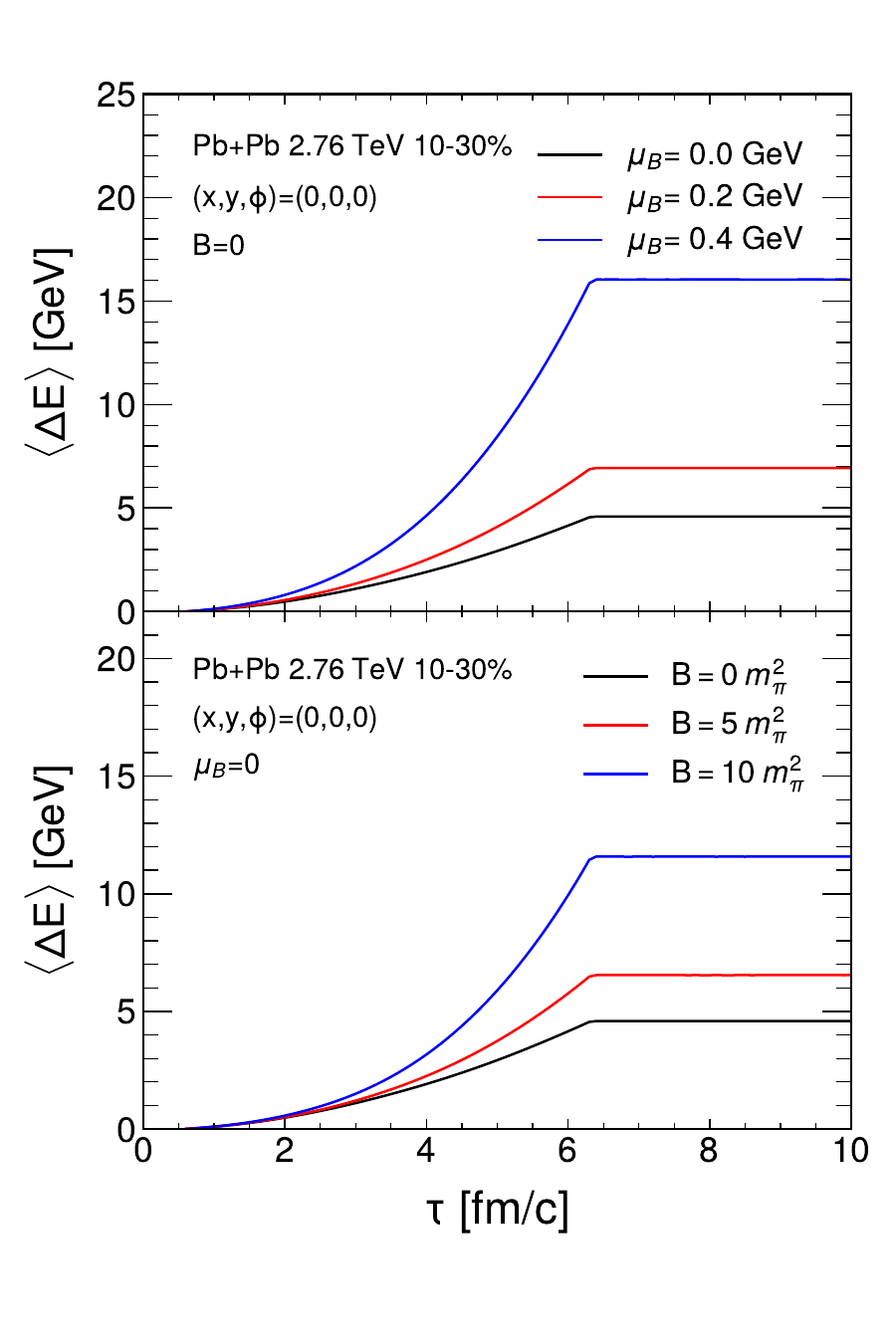}
\includegraphics[width=0.32\textwidth]{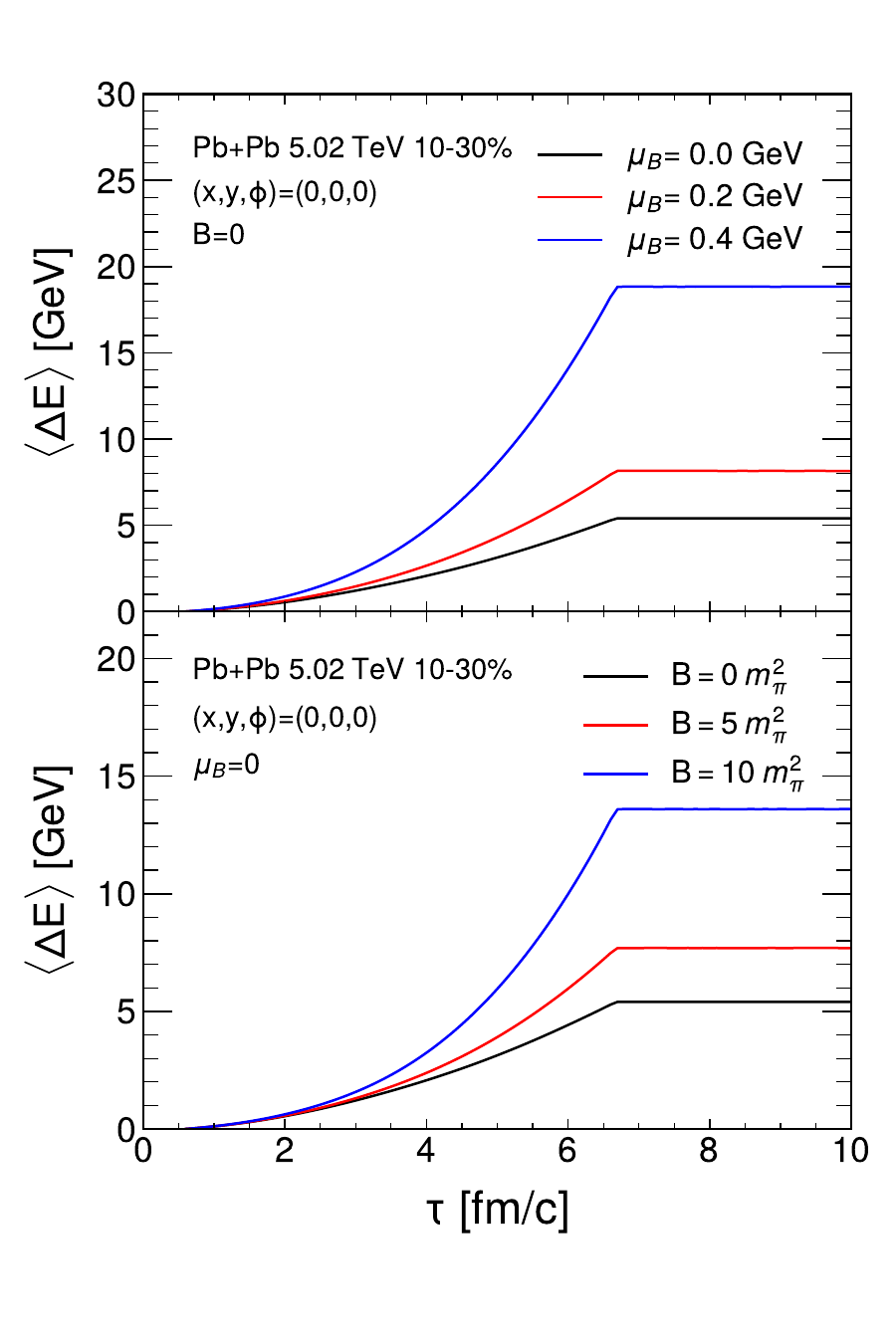}
\end{center}
\caption{The average energy loss dependent on the magnetic field and the chemical potential in 10-20\% Au+Au collisions at $\sqrt{S_{NN}}=200$ GeV, 10-30\% Pb+Pb collisions at $\sqrt{S_{NN}}=2.76, 5.02$ TeV, respectively. Upper panels: The effect of the chemical potential on the average energy loss of jet; Lower panels: The effect of the magnetic field on the average energy loss of jet.}
\label{fig.DeAtTau}
\end{figure*}

\begin{figure}[h]
\begin{center}
\includegraphics[width=0.46\textwidth]{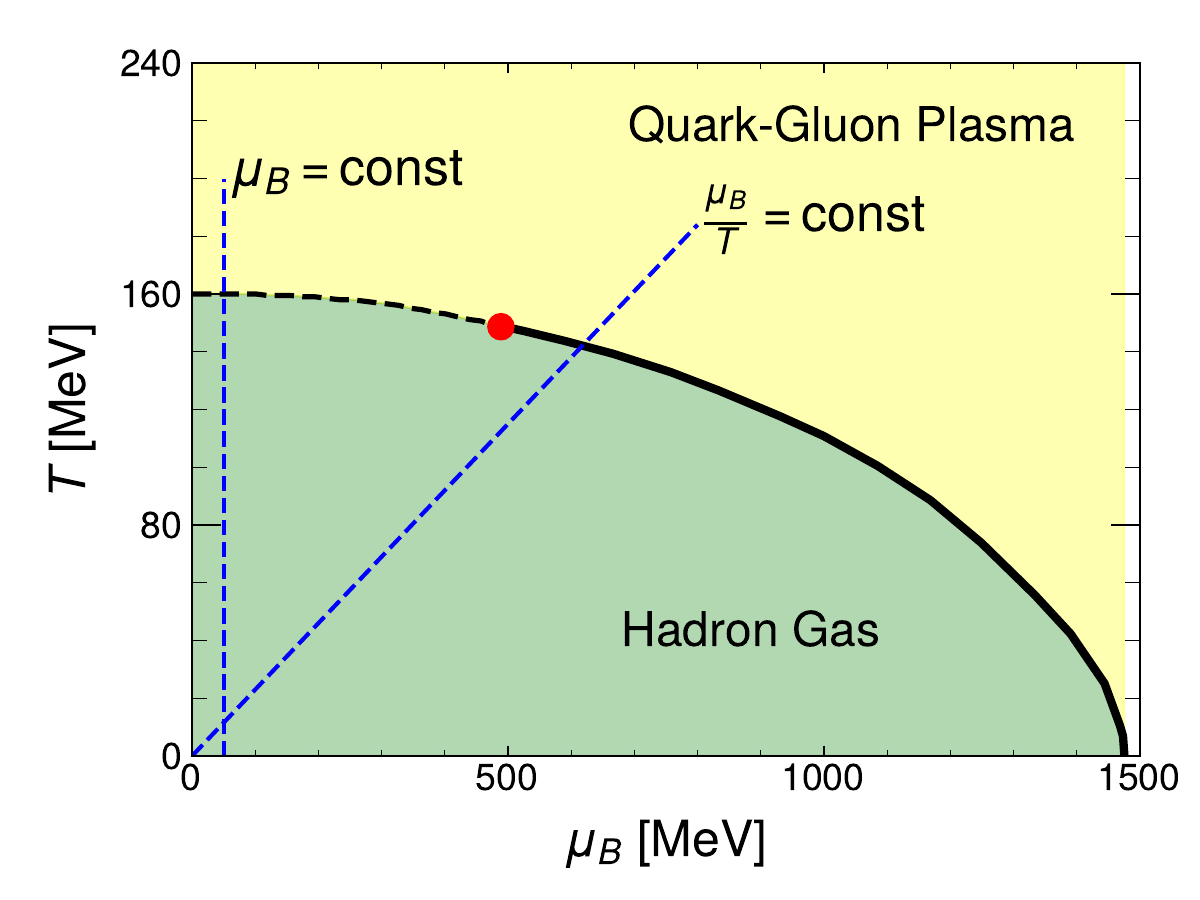} 
\end{center}
\caption{The trajectory of a jet in the phase diagram under the conditions of constant $\mu_{B}$ and constant $\mu_{B}/T$.}
\label{fig:phase}
\end{figure}

\section{Paramtrizing the time-evolution of $B$ and $\mu_B$}

Using this temperature-dependent coupling, we illustrate the effect of $\mu_{B}$ and $B$ in the total energy loss. We calculate the energy loss as a function of $\tau$ for a quark jet that is initially produced at the center of the QGP ($x=y=z=0$) and propagates in the transverse direction ($\eta = 0$) along the direction $\phi = 0$ through the QGP. The energy loss at $\tau$ is the integral of the energy loss rate from $\tau_0$ to $\tau$, $\Delta E(\tau) = \int_{\tau_0}^{\tau} dx dE/dx$. \autoref{fig.DeAtTau} shows the average energy loss dependent on the magnetic field and the chemical potential in 10-20\% Au+Au collisions at $\sqrt{S_{NN}}=200$ GeV, 10-30\% Pb+Pb collisions at $\sqrt{S_{NN}}=2.76, 5.02$ TeV, respectively. 

In principle, the time evolution of the magnetic field and the baryon chemical potential should be obtained from a dynamical model, e.g., magneto-hydrodynamic equations with conserved charge and corresponding equation of state. However, these tools are not yet sophisticated for jet quenching study. Therefore, in this work, we take a very simplistic approach to treat $B$ and $\mu_B$ as background fields and parametrize their time dependence in two extreme cases. 
\begin{align}
\textrm{Scenario I}: &\quad \mu_B = \textrm{const.}, \quad B=\textrm{const.},\\
\textrm{Scenario II}: &\quad \frac{\mu_B}{T} = \textrm{const.}, \quad \frac{B}{T^2}=\textrm{const.}.
\end{align}
The $\mu_B/T$ trajectories of the two cases are illustrated on the phase diagram in figure~\ref{fig:phase}. Even-though these are clearly not the physical case, we hope by tuning the constants in each scenario, it can mimic the physical trajectory from hydrodynamic simulations with finite baryon chemical potential.
As for the magnetic field, scenario II parametrizes a field that decays over time. In figure~\ref{fig.DeAtTau}, we plotted the energy loss with fields turned on in the first scenario. In the upper panel, we set $B$ to 0, and $\mu_{B}$ takes values from 0 to 0.4 GeV. Similarly, we set $\mu_{B}$ to 0 in the bottom panel, and $B$ takes values from 0 to 10 $m_{\pi}^2$. It is clear that the energy loss increases monotonically with $\mu_{B}$ and $B$, which is consistent with Ref~\cite{Zhang:2019jfq, Zhu:2019ujc,Hou:2021own}.


\section{Bayesian inference of $\mu_B$ and $B$
\label{section5}
}
\begin{figure}[h]
\begin{center}
\includegraphics[width=0.40\textwidth]{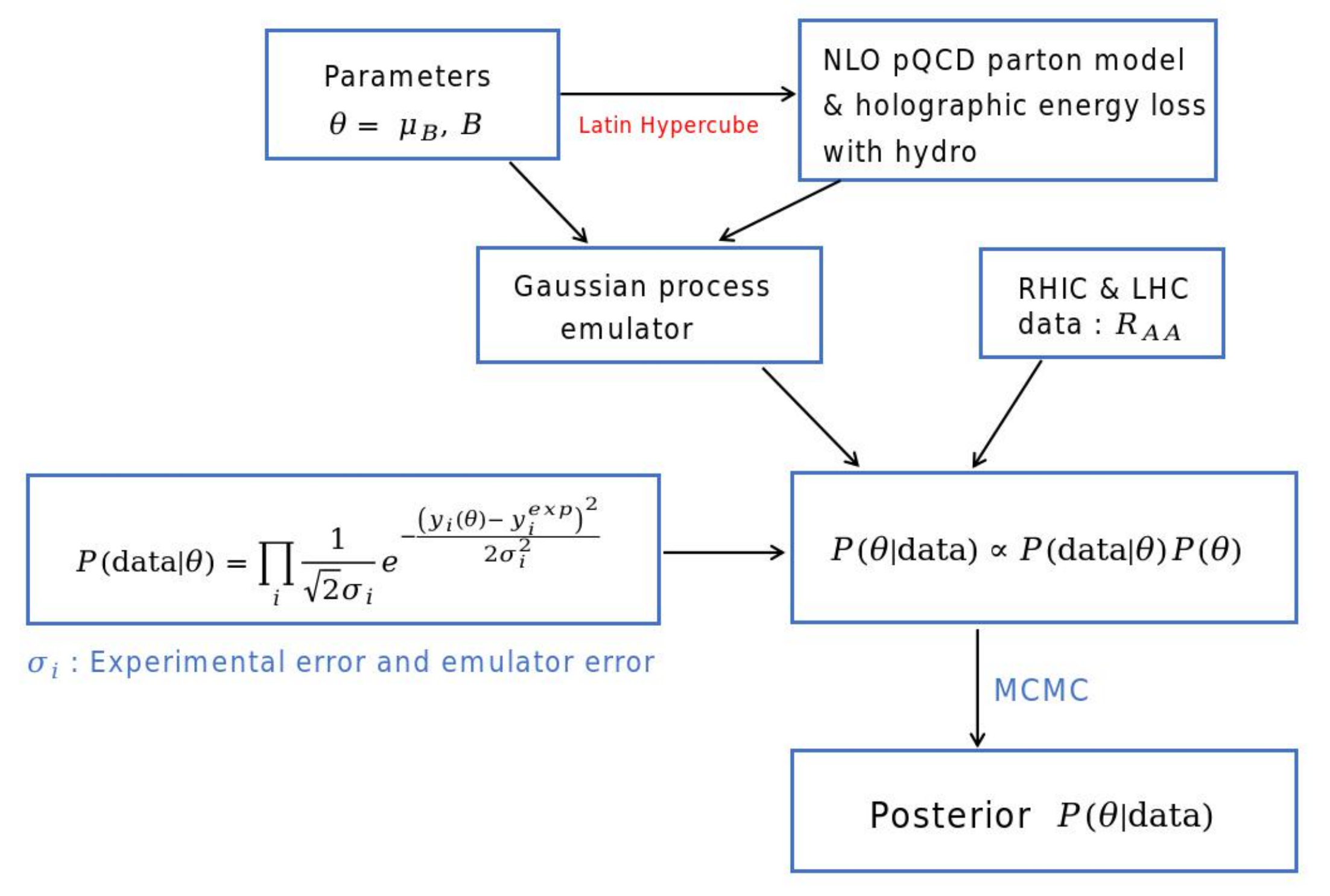} 
\end{center}
\caption{Flowchart of the Bayesian analysis.}
\label{fig:Bayes}
\end{figure}

The method of Bayesian inference for model parameters has achieved significant success in the field of relativistic heavy-ion collisions~\cite{Pratt:2015zsa,Novak:2013bqa,Sangaline:2015isa}. For instance, shear viscosity $\eta/s$ of the QGP medium~\cite{Bernhard:2016tnd, Bernhard:2015hxa}, jet transport coefficient $\hat{q}$~\cite{Xie:2022fak, Xie:2022ght, JETSCAPE:2021ehl}, and the analysis of the jet energy loss~\cite{Wu:2023azi, He:2018gks}. In this work, we utilize the Bayesian inference method to constrain the magnetic field and chemical potential in the energy loss functions of partons traversing the QGP medium, thereby obtaining the variation of the magnetic field and chemical potential in heavy-ion collisions. The Bayesian analysis process is illustrated in figure~\ref{fig:Bayes}.

The set of parameters is denoted by $\theta$. Based on Bayes' theorem:

\begin{align}
\mathit{P}(\theta| \mathrm{data})=\frac{\mathit{P}(\mathrm{data}|\theta)\mathit{P}(\theta)}{P(\mathrm{data})}
\end{align}
$\mathit{P}(\theta| \mathrm{data})$ represents the posterior distribution, which is the conditional probability of the parameters given the observed data. $\mathit{P}(\theta)$ represents the prior distribution, which reflects our initial assumptions or knowledge about the parameters. $\mathit{P}(\mathrm{data} |\theta)$ refers to the likelihood, or the probability of observing the data based on specific assumed parameter values. In our work, $\vec{\theta}=(\mu_{B}, B)$ is a two-dimensional vector, and $\mathit{P}(\mathrm{data}|\theta)$ is assumed to follow a Gaussian distribution. 
\begin{align}
\mathit{P}(\mathrm{data}|\theta)= \frac{1}{(2\pi)^{N/2}|\Sigma|^{1/2}}e^{-\frac{1}{2} [\vec{y}(\theta)-\vec{y}^{\text{exp}}]^T \Sigma^{-1} [\vec{y}(\theta)-\vec{y}^{\text{exp}}] }
\end{align}
Here, $\vec{y}^{\text{exp}}$ is the vectorized experimental data. $\vec{y}(\theta)$ is the vectorized model output including all observables. This model evaluation is surrogated by a Gaussian emulator~\cite{williams2006gaussian, 10.1214/ss/1177012413} that is trained using at 300 sets of input parameters with full-model calculations. $\Sigma$ is the covariance matrix that have included both experimental uncertainty and Gaussian emulator uncertainty.

In our following calculations, we set two cases of constant magnetic field and chemical potential, as shown in figure~\ref{fig:phase}.
We firstly assume that both the magnetic field and chemical potential remain constant throughout the entire heavy-ion collision process without changing over time. Secondly we assume that the dimensionless observables remain constant during the collision process, which are the ratio of the magnetic field over temperature $T$ squared and the ratio of the chemical potential over temperature $T$. 
Given the complexity of jet behavior as it moves through the medium in heavy-ion collisions, we focus solely on the motion of the jet under these two limiting conditions.
Starting from the prior, we set a uniform prior distribution: $\mu_{B}\in$ (0, 0.3) $\mathrm{GeV}$ and $B\in$ (0, 15) $m_{\pi}^{2}$ in the first assumption, and $\mu_{B}/T\in$ (0, 3) and $B/T^{2}\in$ (0, 10) in the second assumption. Within this prior range, we use Latin hypercube sampling to obtain 300 design points. These 300 sets of parameters were then input into the $\mathrm{NLO}$ $\mathrm{pQCD}$ parton model to calculate the nuclear modification factor $R_{AA}$ for hadrons in different-centrality Au+Au collisions at 200 GeV and in different-centrality Pb+Pb collisions at 2.76 or 5.02 TeV. We then trained these 300 design points and their corresponding model outputs into a Gaussian emulator for use in subsequent Bayesian inference.

Next, we use Markov chain
Monte Carlo (MCMC)~\cite{Foreman-Mackey:2012any, Goodman:2010dyf} for Bayesian calibration. MCMC is based on the Metropolis-Hastings (M-H) algorithm for sampling posterior probabilities. This algorithm generates samples by performing a random walk in the parameter space and accepting or rejecting each step based on the posterior probability, in order to extract the probability distribution of parameter space $\theta$.

\section{Extracting magnetic field and chemical potential
\label{sec:Result}
}

\begin{figure*}
\begin{center}
\includegraphics[width=0.33\textwidth]{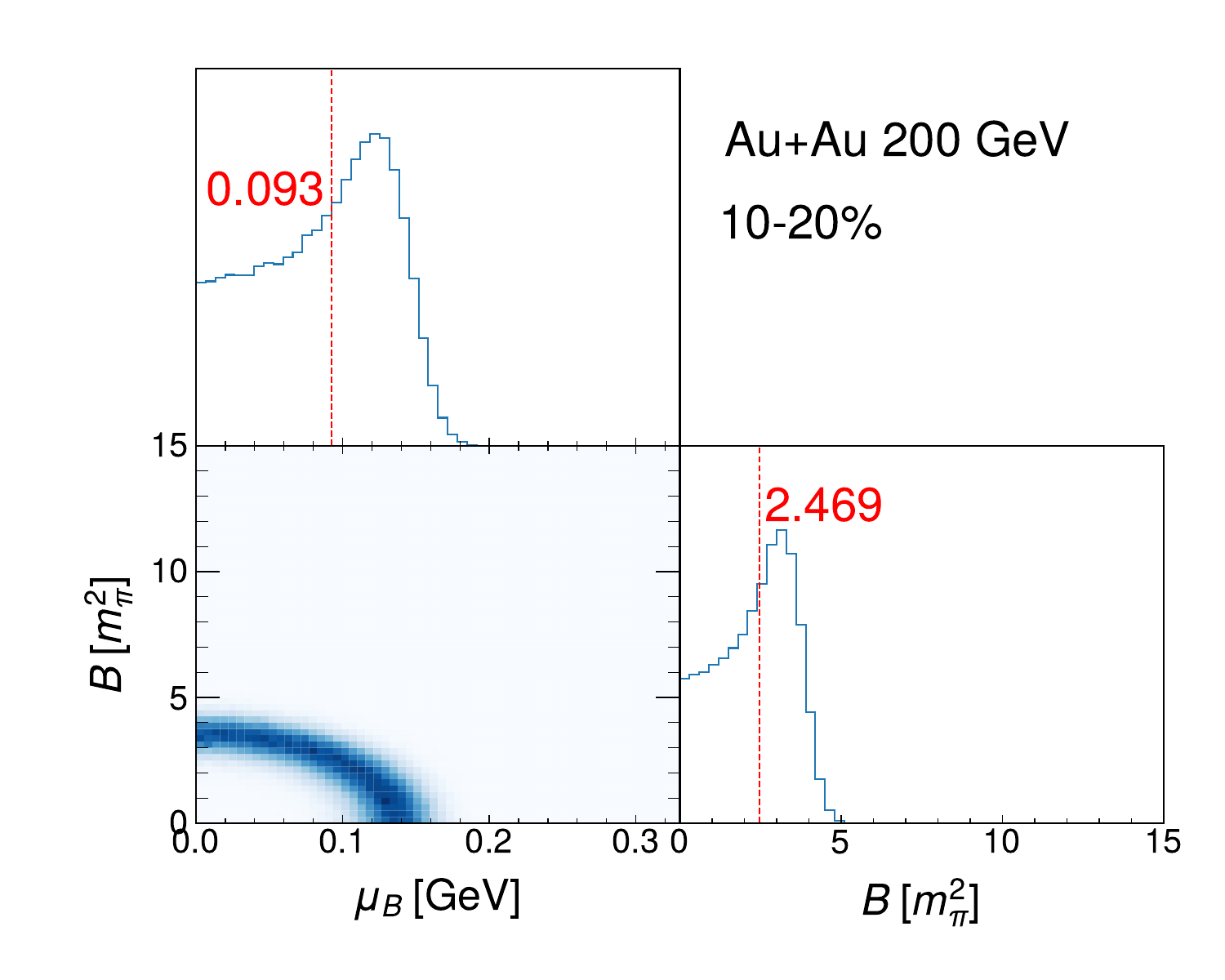}
\includegraphics[width=0.33\textwidth]{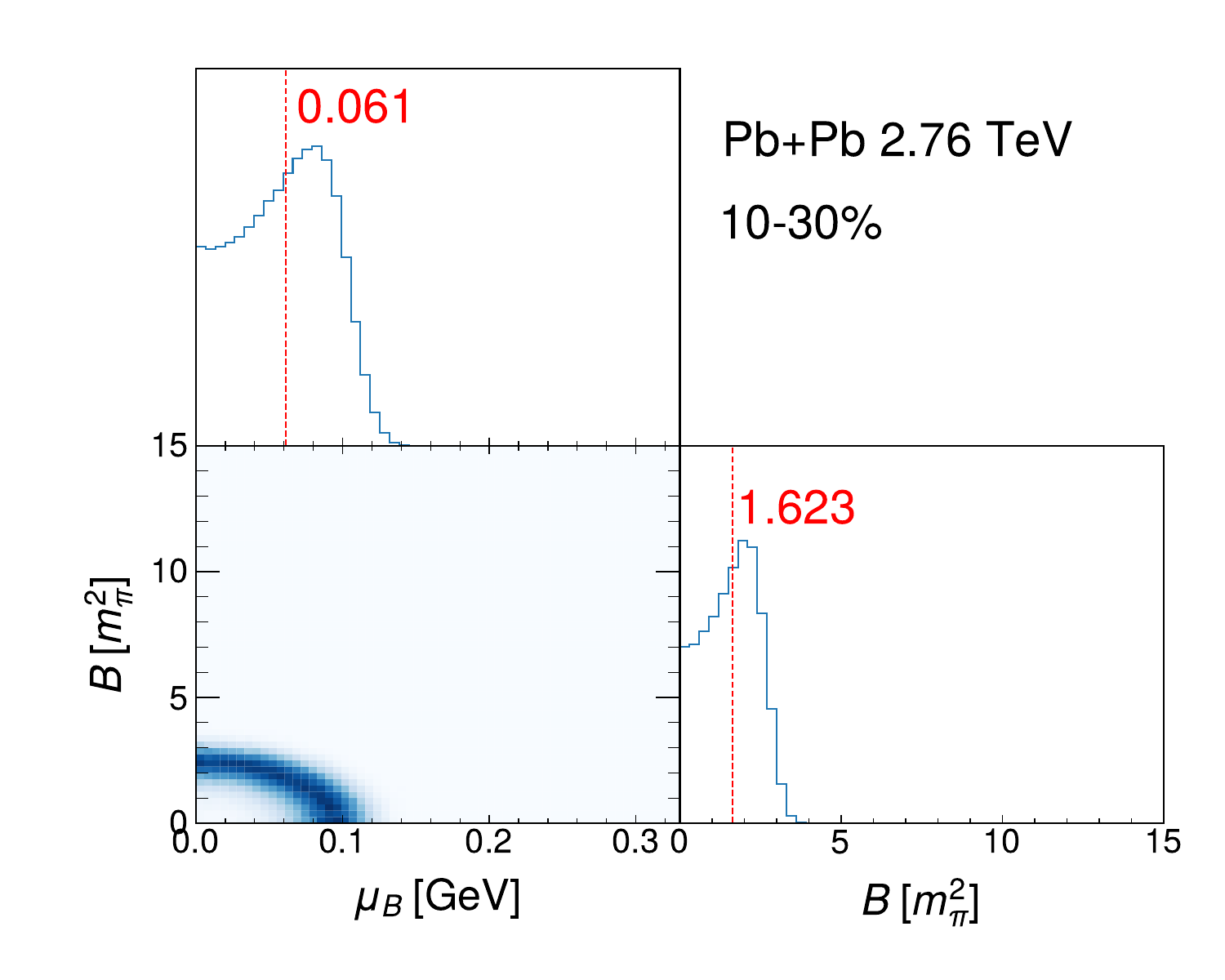}
\includegraphics[width=0.32\textwidth]{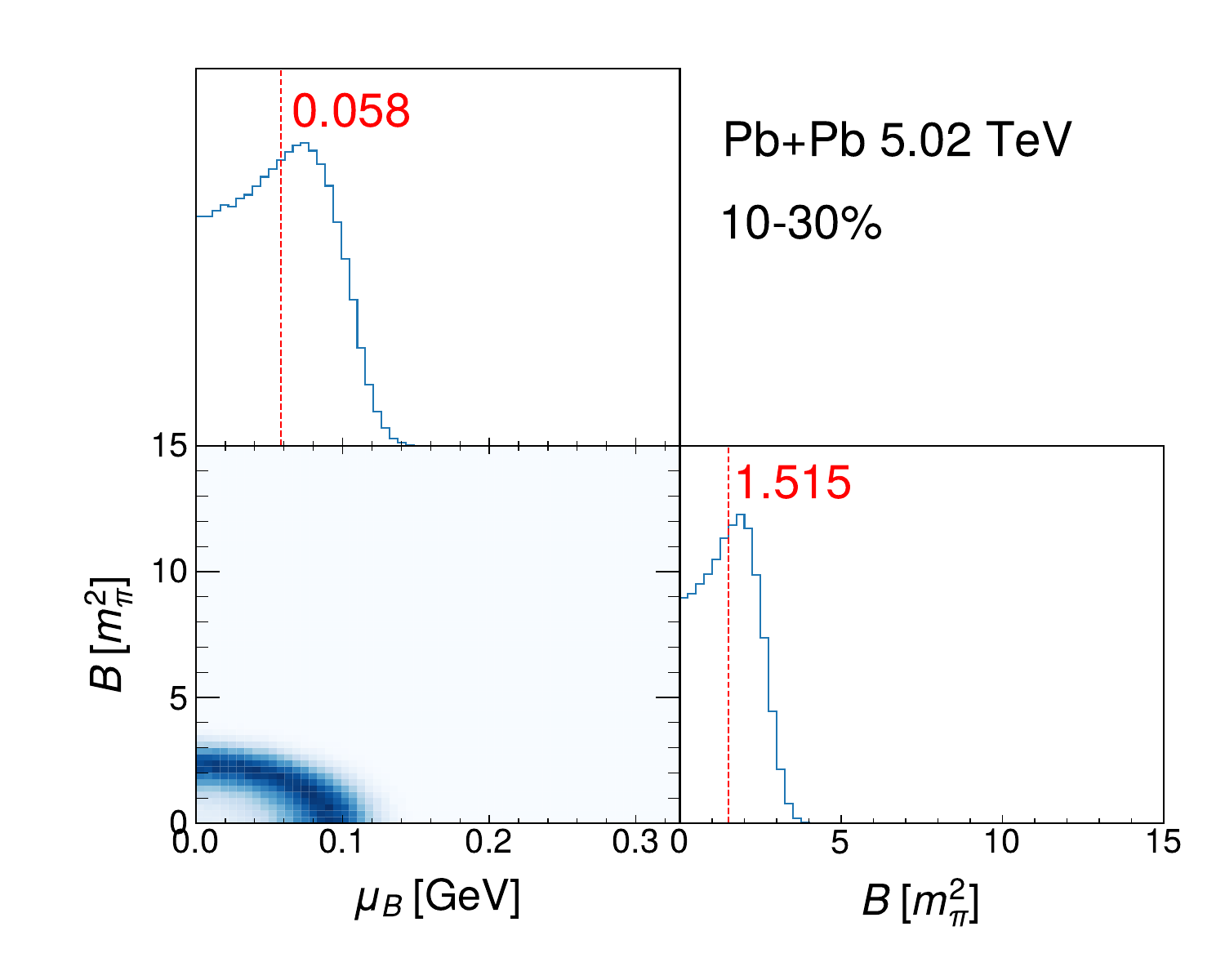}\\
\vspace{0.3cm}
\includegraphics[width=0.33\textwidth]{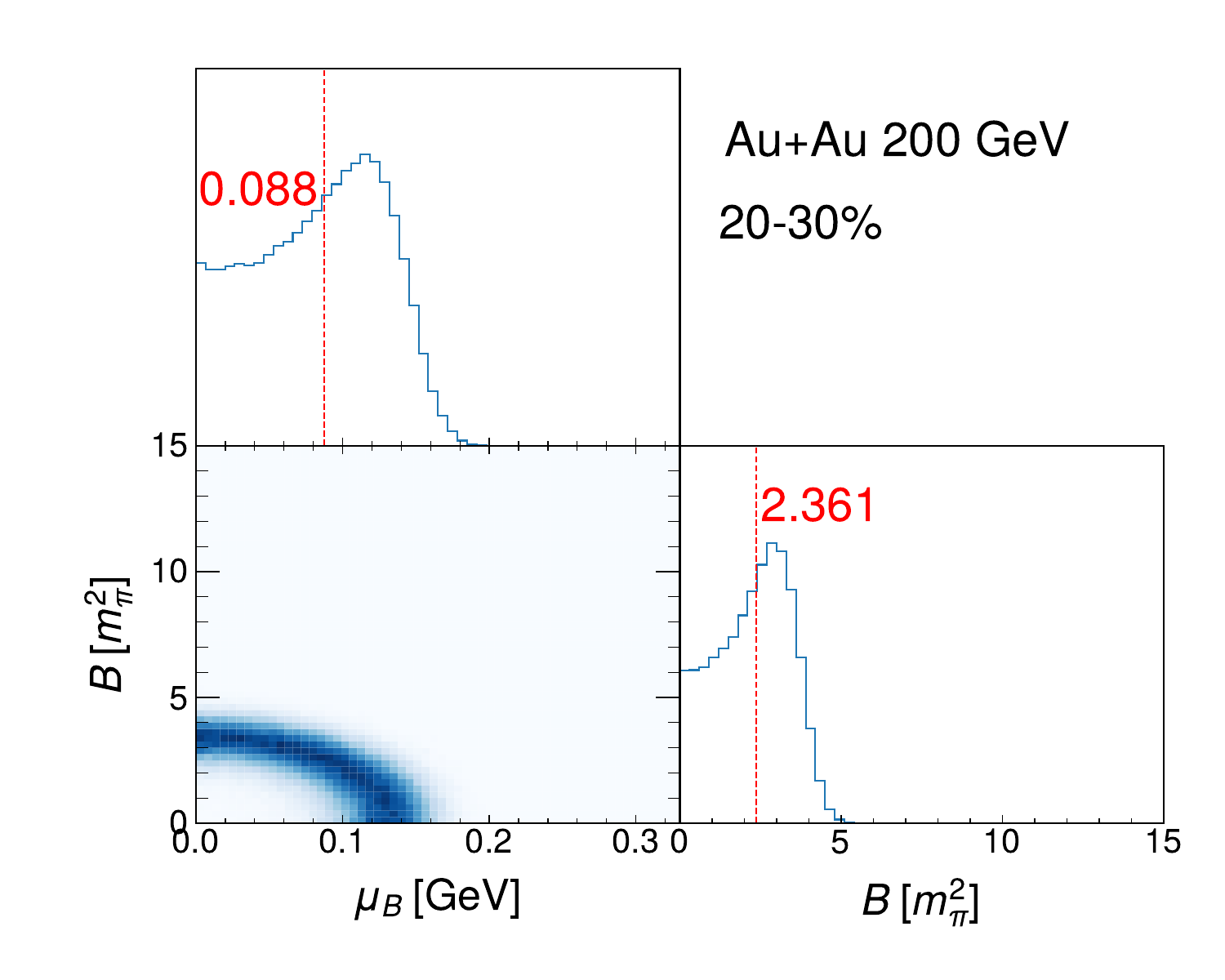}
\includegraphics[width=0.33\textwidth]{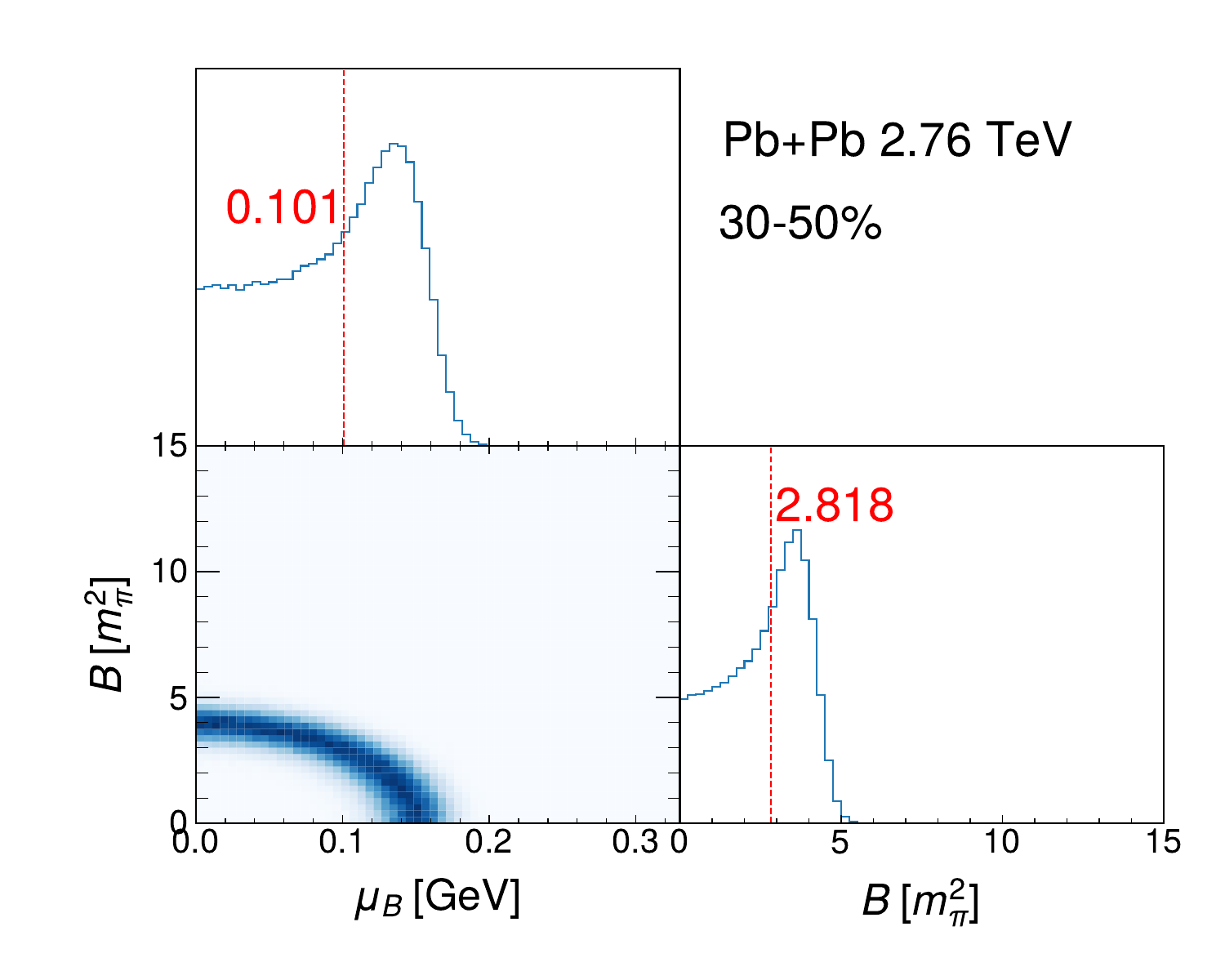}
\includegraphics[width=0.33\textwidth]{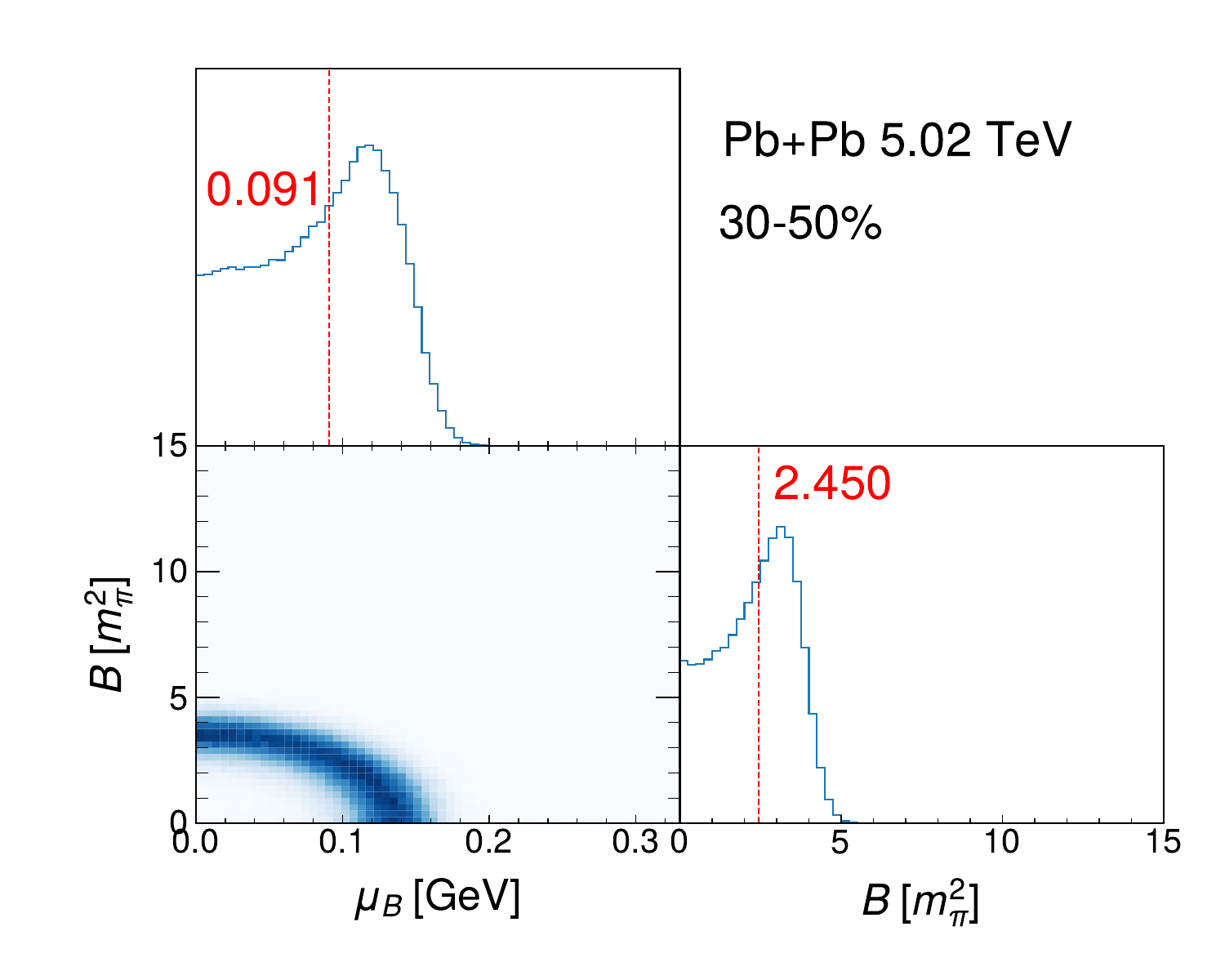}\\
\vspace{0.3cm}
\includegraphics[width=0.33\textwidth]{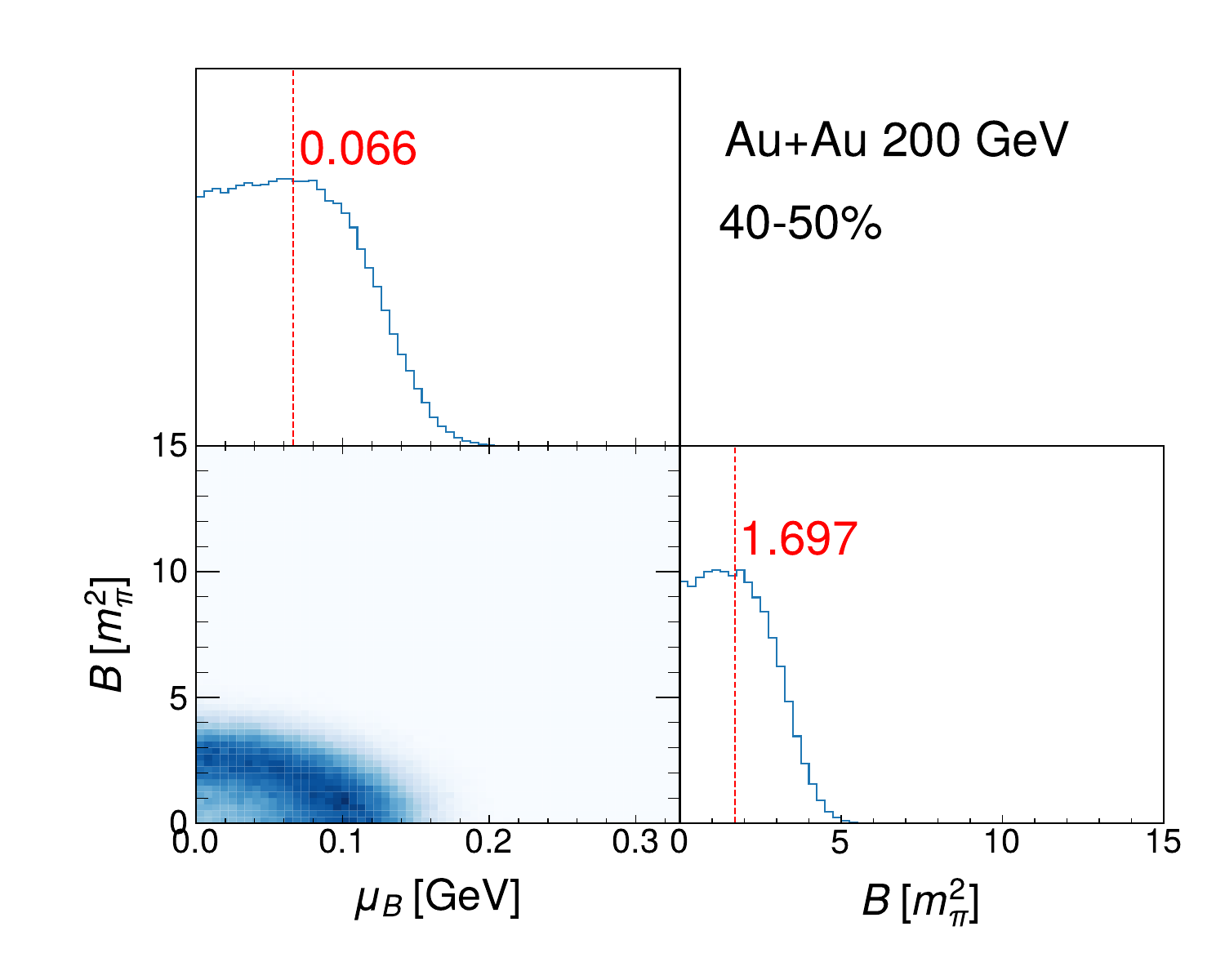}
\includegraphics[width=0.33\textwidth]{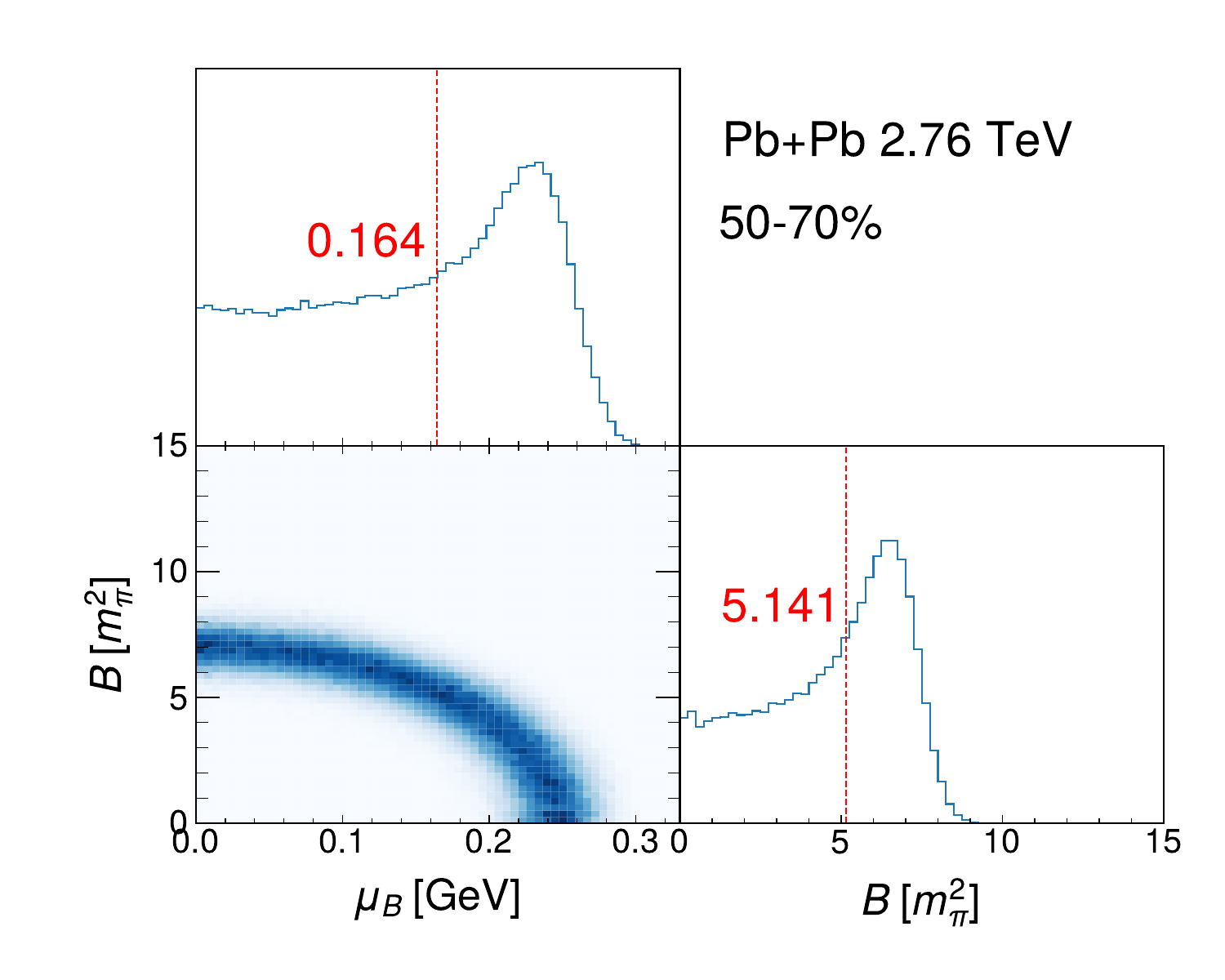}
\includegraphics[width=0.33\textwidth]{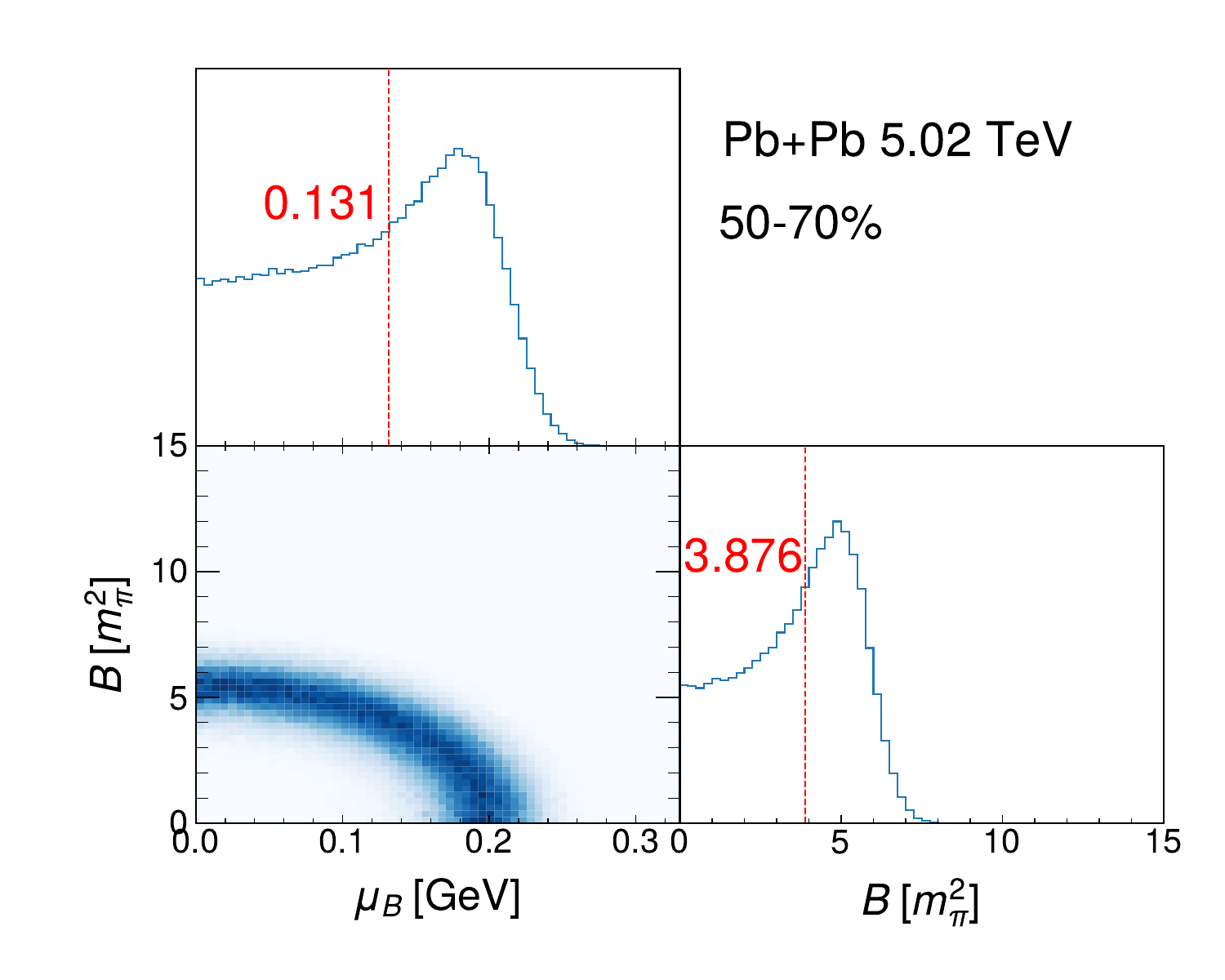}
\end{center}
\caption{The Bayesian inference of the magnetic field and chemical potential (diagonal panels) and their correlations (off-diagonal panels). Left column: for 10-20\%, 20-30\%, and 40-50\% Au+Au collisions at $\sqrt{s_{NN}}$ = 200 GeV. Center column: for 10-30\%, 30-50\%, and 50-70\% Pb+Pb collisions at $\sqrt{s_{NN}}$ = 2.76 TeV. Right column: for 10-30\%, 30-50\%, and 50-70\% Pb+Pb collisions at $\sqrt{s_{NN}}$ = 5.02 TeV. The data are from PHENIX and CMS on the charged-hadron $R_{AA}$~\cite{PHENIX:2012jha, CMS:2012aa, CMS:2016xef}.}
\label{fig:bayes1}
\end{figure*}
\begin{figure*}
\begin{center}
\includegraphics[width=0.33\textwidth]{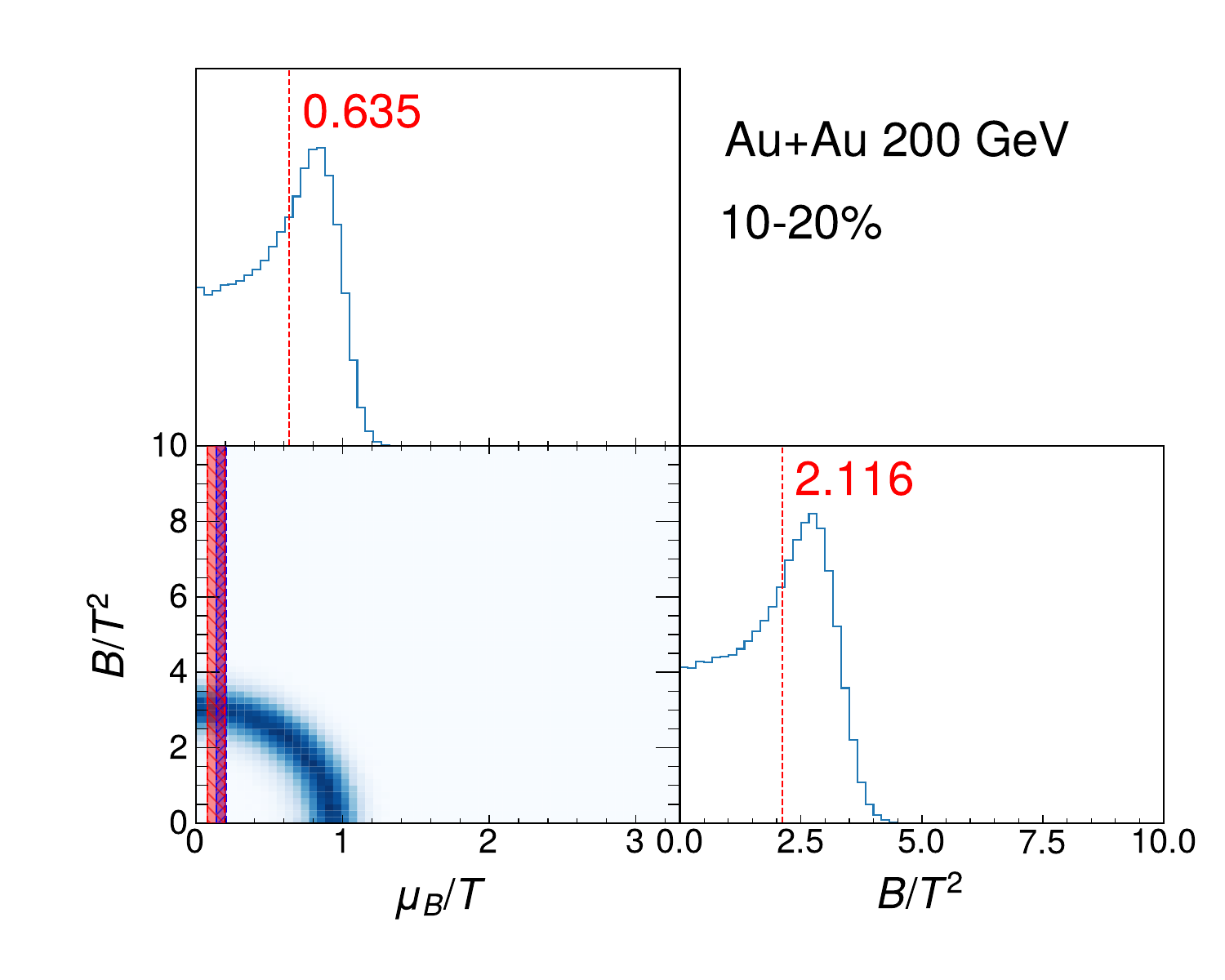}
\includegraphics[width=0.33\textwidth]{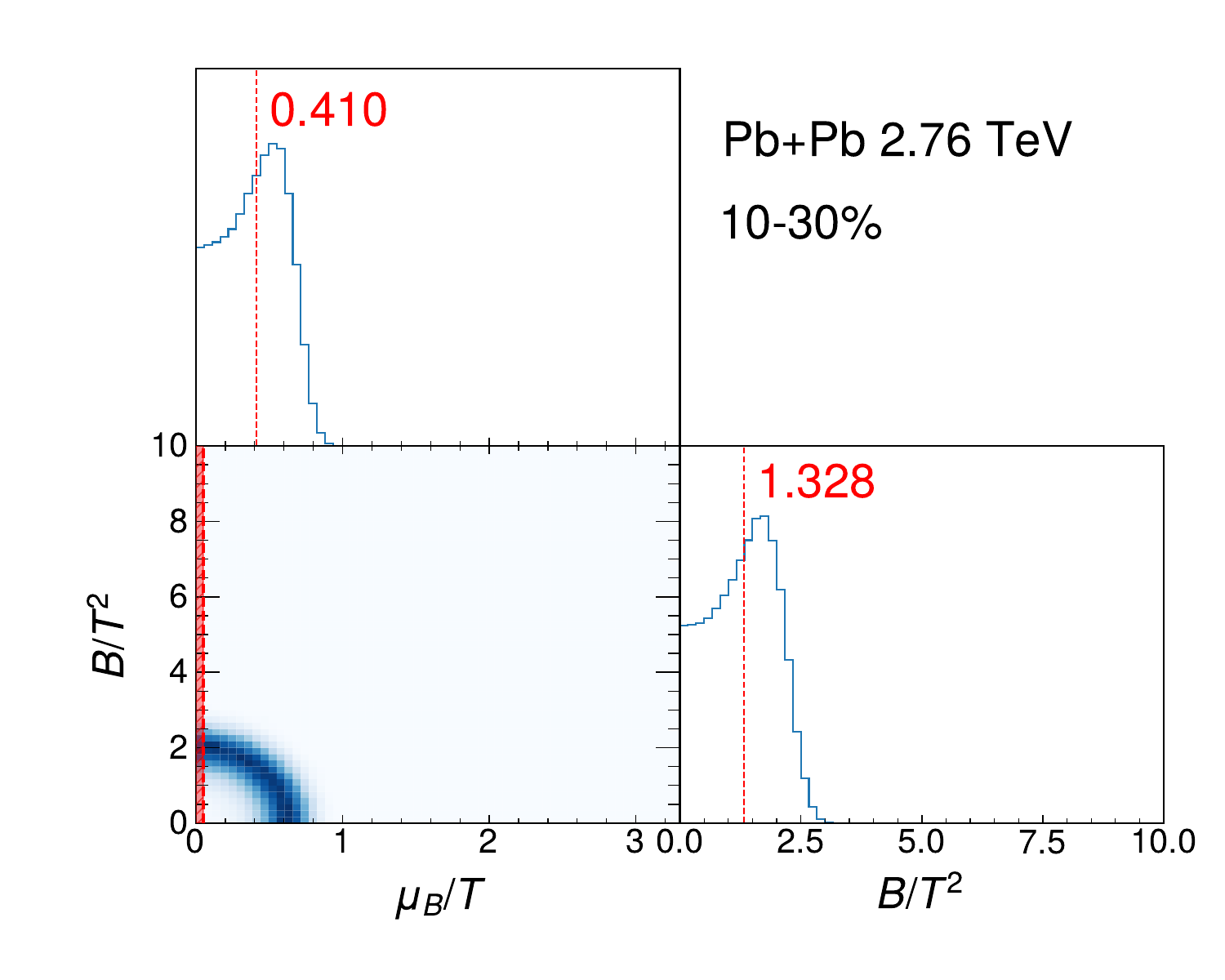}
\includegraphics[width=0.33\textwidth]{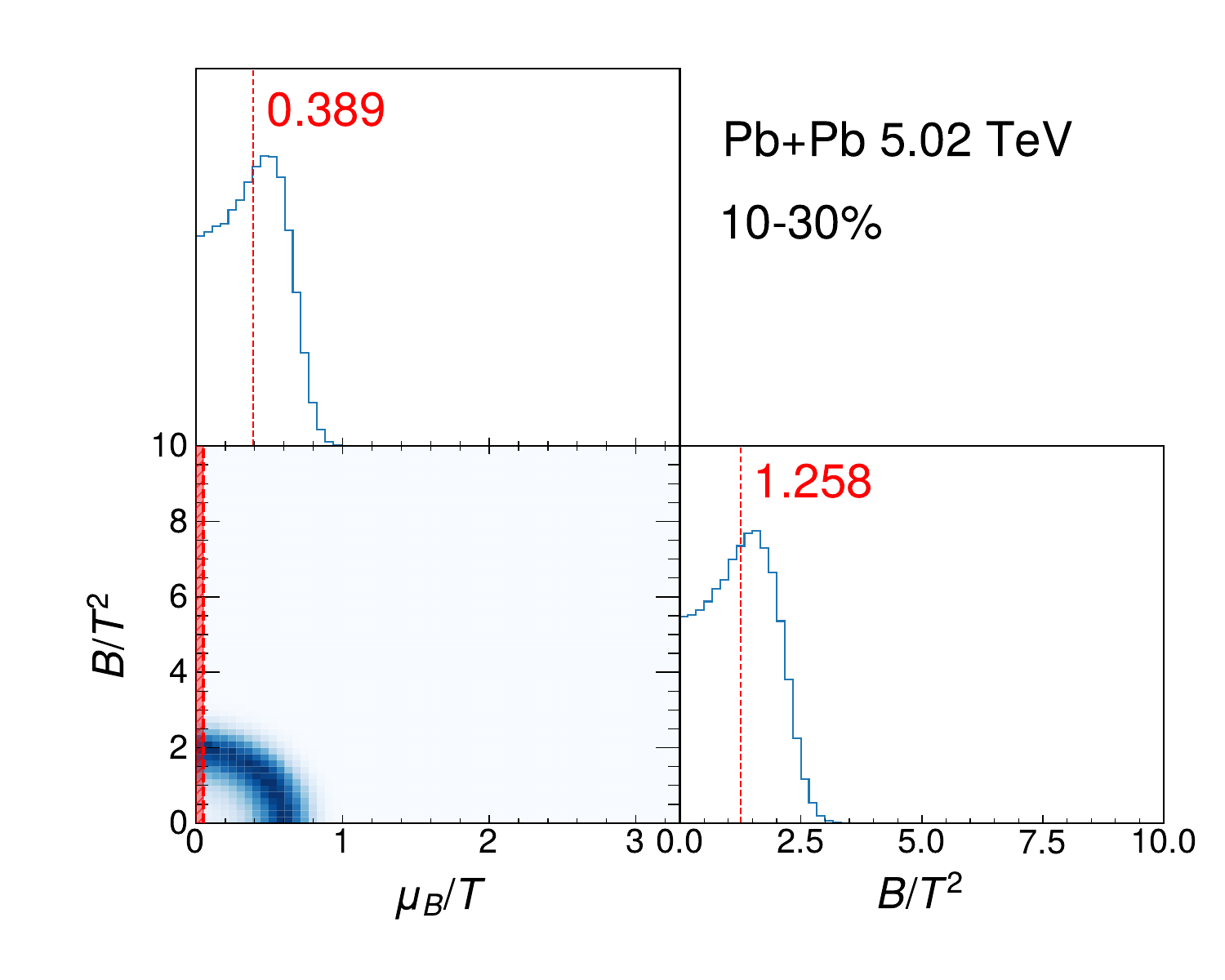}\\
\vspace{0.3cm}
\includegraphics[width=0.33\textwidth]{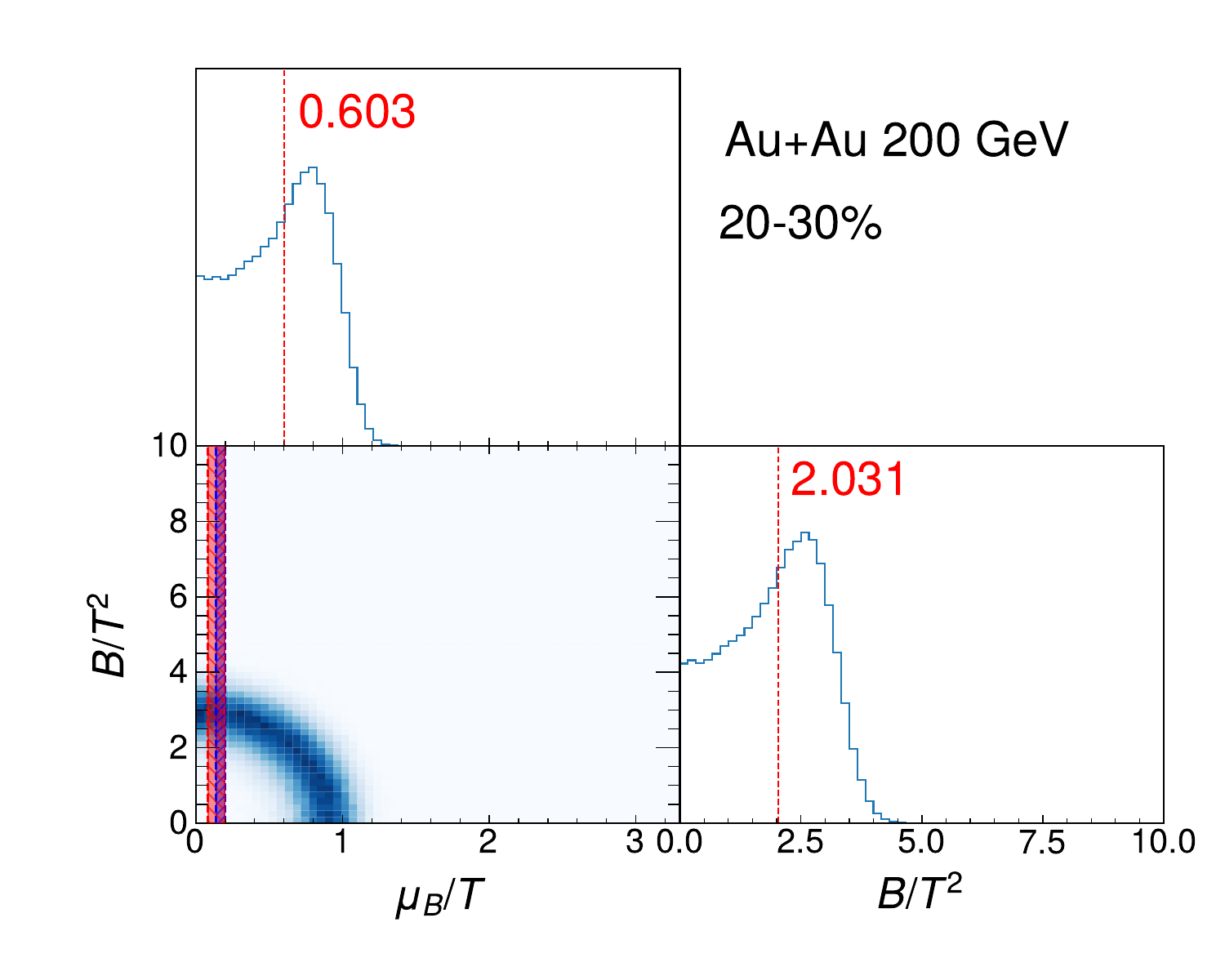}
\includegraphics[width=0.33\textwidth]{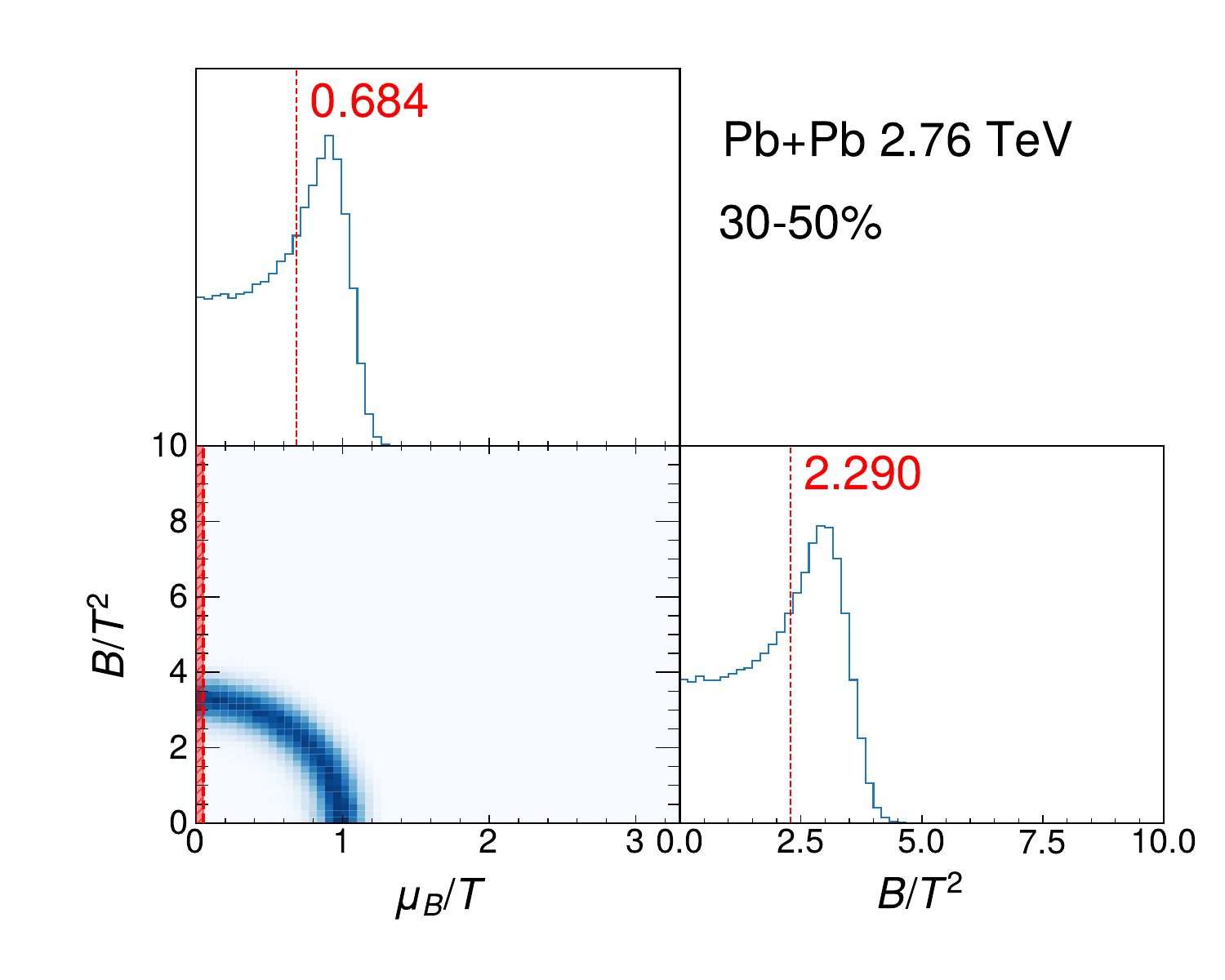}
\includegraphics[width=0.33\textwidth]{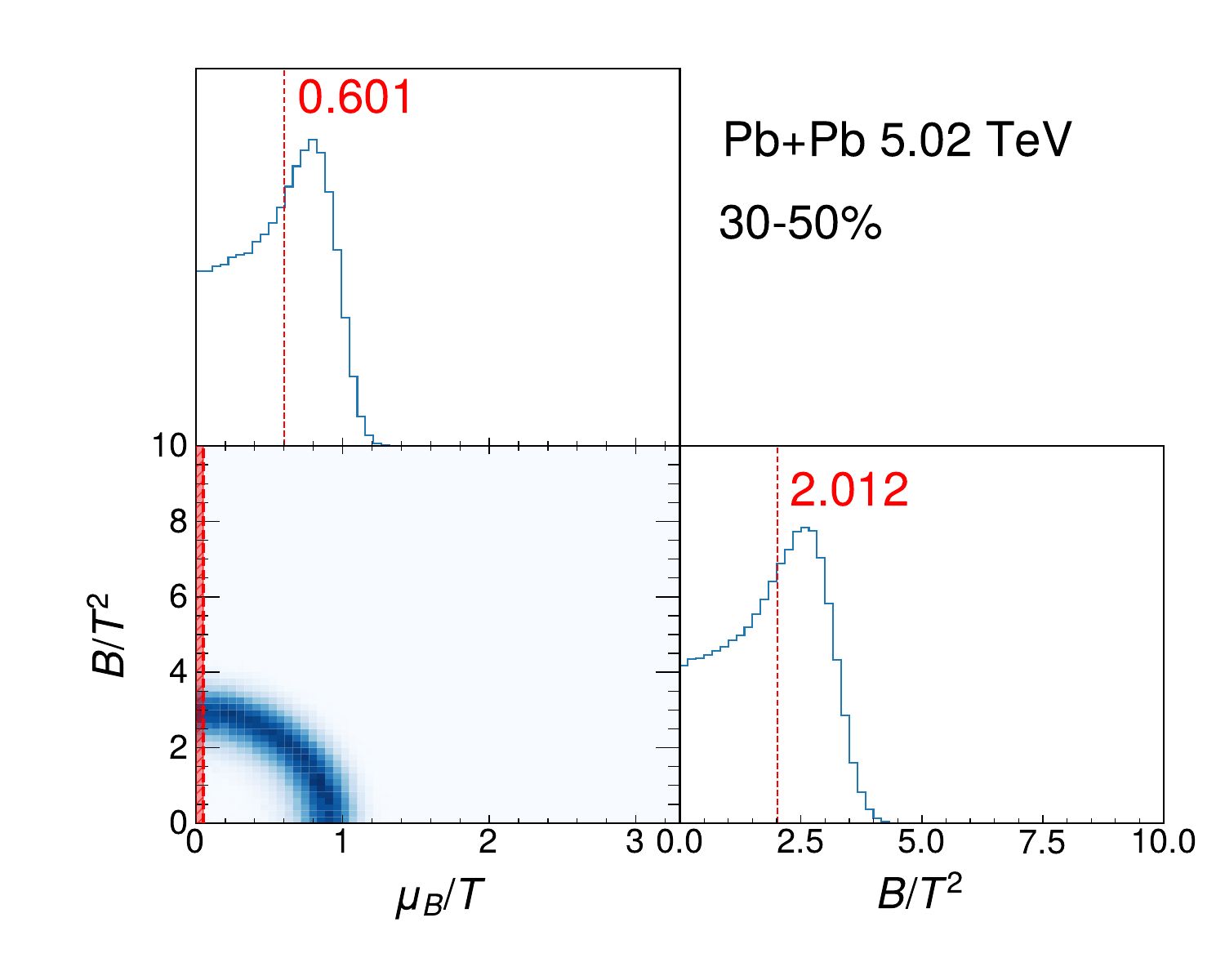}\\
\vspace{0.3cm}
\includegraphics[width=0.33\textwidth]{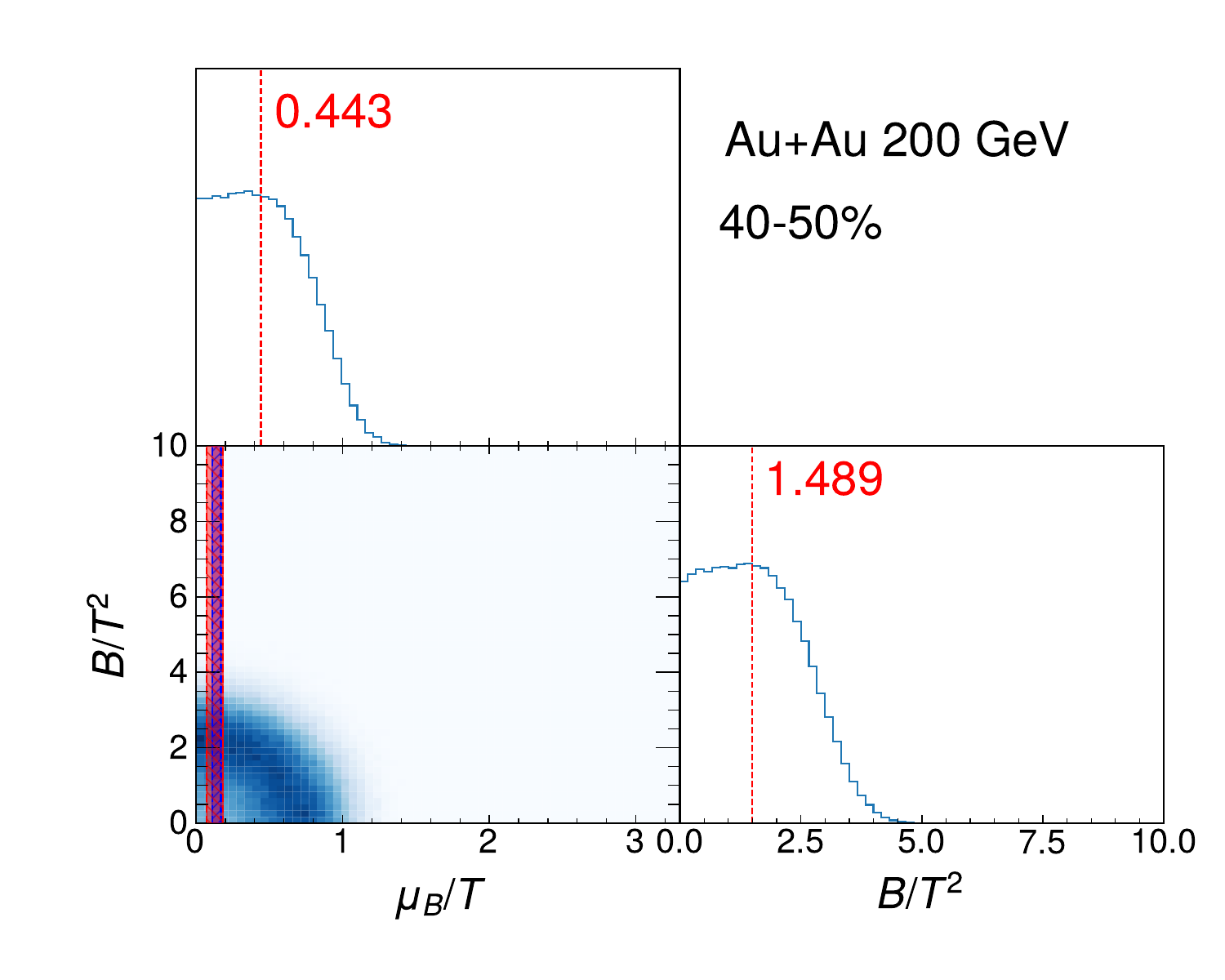}
\includegraphics[width=0.33\textwidth]{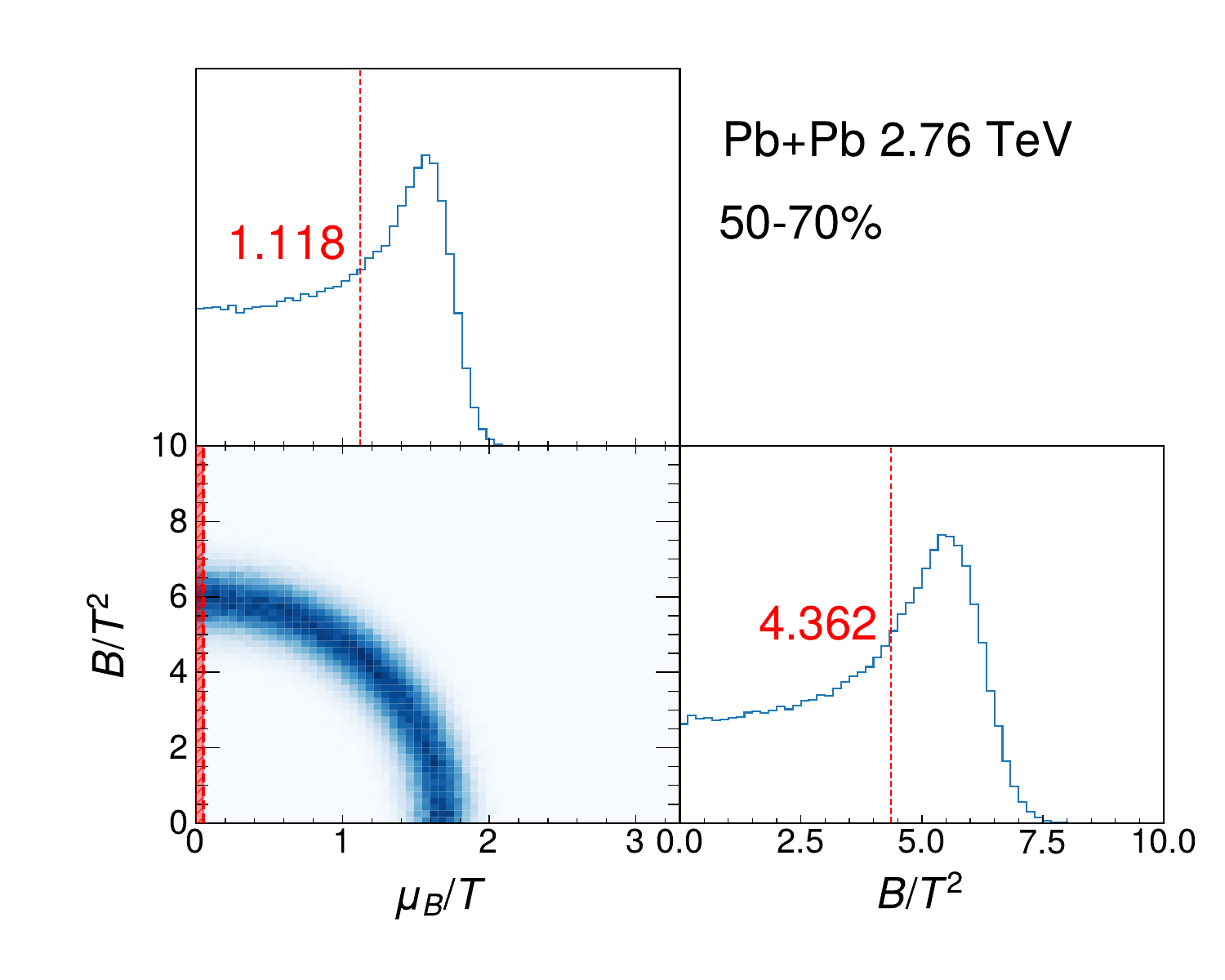}
\includegraphics[width=0.33\textwidth]{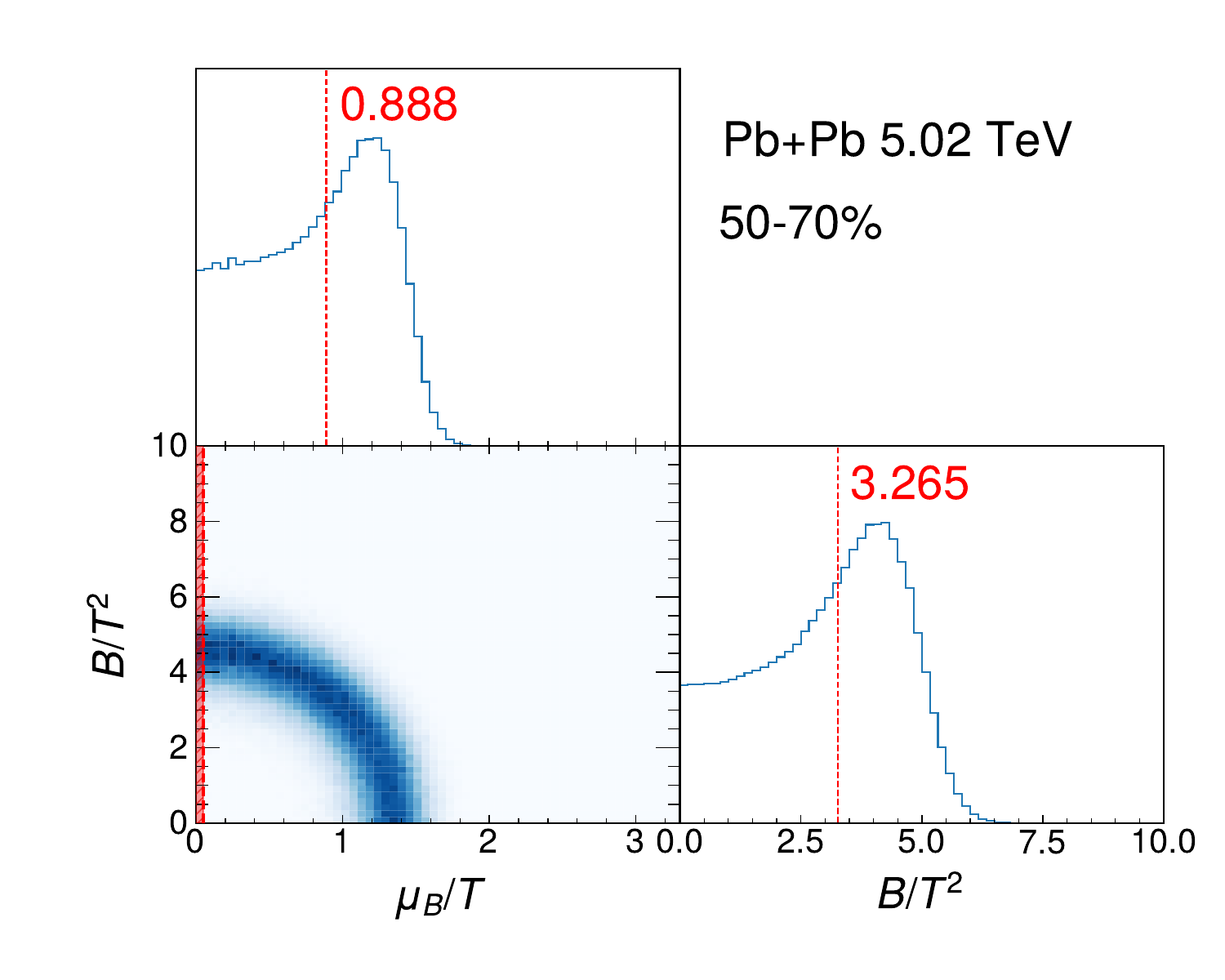}
\end{center}
\caption{The posterior distributions of the scaled magnetic field and scaled chemical potential (diagonal panels) and their correlations (off-diagonal panels). Left column: 10-20\%, 20-30\%, and 40-50\% Au+Au collisions at $\sqrt{s_{NN}}$ = 200 GeV. Center column: 10-30\%, 30-50\%, and 50-70\% Pb+Pb collisions at $\sqrt{s_{NN}}$ = 2.76 TeV. Right column: 10-30\%, 30-50\%, and 50-70\% Pb+Pb collisions at $\sqrt{s_{NN}}$ = 5.02 TeV. Wth experimental data sourced from PHENIX and CMS on the $R_{AA}$ of charged light hadrons~\cite{PHENIX:2012jha, CMS:2012aa, CMS:2016xef}. In Au+Au collisions, the red and blue rectangular shaded regions correspond to the experimental ranges of the chemical freeze-out temperature to baryon chemical potential ratio for the 10-20\%, 20-30\%, and 40-60\% centrality, respectively ~\cite{STAR:2017sal}. Since experimental data are not available for the 40-50\% centrality, we compare our 40-50\% centrality results with the experimental results from the 40-60\% centrality in our calculations. In Pb+Pb collisions at LHC, the ratio of chemical freeze-out temperature to baryon chemical potential approaches nearly zero~\cite{Lysenko:2024hqp}, as indicated by the red rectangular shaded region.}
\label{fig:bayes2}
\end{figure*}

Based on the Bayesian inference framework from last section, let's extract the distributions of magnetic field and chemical potential from the nuclear modification factor $R_{AA}$ of data~\cite{PHENIX:2012jha, CMS:2012aa, CMS:2016xef} for the hadron production across different-centrality collision systems and different collision energies. These data are for 10-20\%, 20-30\% and 40-50\% Au+Au collisions at $\sqrt{S_{NN}}=200$ GeV, and 10-30\%, 30-50\% and 50-70\% Pb+Pb collisions at $\sqrt{S_{NN}}=2.76$ or 5.02 TeV, respectively. We will first begin with the constant magnetic field and chemical potential and then the scaled magnetic field and chemical potential.

Shown in figure~\ref{fig:bayes1} are the 
Bayesian inference of the magnetic field and chemical potential (diagonal panels) and their correlations (off-diagonal panels) in different-centrality A+A collisions at 0.2, 2.76 and 5.02 TeV. These figures present the posterior probability distributions of the magnetic field and chemical potential obtained by fitting experimental data for nuclear modification factors. The diagonal panels show the marginal distributions of the magnetic field and chemical potential (The red dashed line indicates the position of the median), while the off-diagonal panels display their joint distributions. The left column corresponds to $\sqrt{s_{NN}}$ = 200 GeV with centralities of 10-20\%, 20-30\%, and 40-50\% in Au-Au collisions; the middle column corresponds to
 $\sqrt{s_{NN}}$ = 2.76 TeV with centralities of 10-30\%, 30-50\%, and 50-70\% in Pb-Pb collisions; and the right column corresponds to $\sqrt{s_{NN}}$ = 5.02 TeV with the same centralities in Pb-Pb collisions.
 
The results demonstrate a good alignment between the parameters of the magnetic field and chemical potential with the experimental data. A distinct negative correlation is observable between the magnetic field and chemical potential. This correlation stems from the fact that both parameters augment jet energy loss in the same direction, meaning an increase in one can be counterbalanced by a decrease in the other to uphold consistency. Further investigation shows that, at a constant collision centrality, the chemical potential diminishes as the collision energy escalates, which is in accordance with the findings in Ref.~\cite{Andronic:2017pug}. Given the negative correlation between the chemical potential and magnetic field, the magnetic field also diminishes as the collision energy increases, aligning with the conclusion of Ref.~\cite{Deng:2012pc}. At a fixed collision energy, as the collision geometry becomes more eccentric (i.e., centrality rises), both the magnetic field and chemical potential intensify.

The $\mu_{B}$ and $B$ obtained under the constant assumption represent an overall effect. However, the magnetic field strength decays rapidly at the early stage, then decays inversely proportional to the time once the QGP is formed and responds to the electromagnetic field~\cite{Deng:2012pc}. Thus, one can assume that $\mu_{B}$ and $B$ follow the same pattern of time evolution as $T$.
Shown in figure~\ref{fig:bayes2} are the scaled magnetic fields ($B/T^2$) and the scaled chemical potentials ($\mu_{B}/T$). The calculation methodology is similar to that in figure~\ref{fig:bayes1}.
The results show that, at the same collision centrality, the scaled magnetic field and the scaled chemical potential decrease as the collision energy increases. Conversely, at the same collision energy, as the degree of collision eccentricity rises (i.e., as centrality increases), both the scaled magnetic field and scaled chemical potential exhibit a rising trend. These findings provide further insight into the systematic behavior of the magnetic field and chemical potential in high-energy nuclear collisions. 
In Au+Au collisions, the experimental ranges of the ratio of chemical freeze-out temperature to baryon chemical potential are provided for the centralities of 10-20\% ,20-30\% and 40-60\%. The red rectangular shaded region represents the results obtained using the GCEY method, while the blue rectangular shaded region represents the results obtained using the GCER method. Both methods are sourced from Ref.~\cite{STAR:2017sal}, Since the experimental literature does not provide data for 40-50\% centrality, we will use the experimental results for 40-60\% centrality as an approximation and compare them with our calculated results for 40-50\% centrality.
However, in Pb+Pb collisions at LHC, the ratio of chemical freeze-out temperature to baryon chemical potential approaches zero~\cite{Lysenko:2024hqp}, as shown by the red rectangular shaded regio.

Using the above extraction values for the magnetic field and the chemical potential as well as the scaled magnetic field and the scaled chemical potential, we show in the Appendix for the prior and the posterior nuclear modification factors $R_{AA}$ compared with data in different-centrality A+A collisions at 0.2, 2.76 and 5.02 TeV, respectively. Our posterior results fit data well.

\section{Summary and outlook
\label{sec:outlook}
}

In relativistic heavy-ion collisions, the jet quenching effect is an important phenomenon that provides a unique experimental window for studying the microscopic properties and dynamical behavior of the QGP. Jet quenching describes the energy loss of high energy jets as they traverse the QGP, caused by strong interactions with the medium. This effect can be indirectly observed through the experimentally measured transverse momentum nuclear modification factor $R_{AA}$. However, the specific impact of the magnetic field and chemical potential in the QGP medium on jet energy loss has not yet been thoroughly investigated.
Inspired by this scientific question, we derived a jet energy loss formula incorporating the effects of the magnetic field and chemical potential based on the AdS/CFT correspondence, which describes the energy attenuation of partons moving through the QGP medium. To validate the applicability of the theoretical model and to reveal the influence of the magnetic field and chemical potential on jet quenching, we employed Bayesian inference to systematically compare the theoretical predictions with the experimentally measured $R_{AA}$ data. Using this approach, we simultaneously extracted, for the first time, the distributions of the magnetic field and chemical potential in the QGP under different collision energies and centrality.

Our findings reveal the following key patterns:
(i) Enhancement of Jet Energy Loss by the Magnetic Field and Chemical Potential: The presence of the magnetic field and chemical potential significantly enhances the jet energy loss effect, indicating that these properties of the QGP medium play a crucial role in the strong interactions between partons and the medium; 
(ii) Impact of Collision Centrality: As the collisions tend to become more eccentric, the strengths of the magnetic field and chemical potential increase significantly. This may reflect the more pronounced local charge asymmetry generated in the collisions;
(iii) Impact of Collision Energy: With increasing collision energy, both the magnetic field and chemical potential exhibit a decreasing trend. This phenomenon may be related to the expansiveness of the QGP and the changes in initial conditions at higher energies;
(iv) Theoreticl Caluclations Show that Magnetic Field and Chemical Potential affects the energy loss in a highly correlated manner: extracting the energy loss rate alone cannot provide a unique sensitivy to only one of the two effects.

These research findings not only enrich our theoretical understanding of the effects of the magnetic field and chemical potential on jet energy loss but also provide a new direction for experimentally exploring the physical properties of the QGP using $R_{AA}$ data. 

\section{Acknowledgments}
\label{sec:acknowledgements}

We extend our gratitude to Prof. Defu Hou and Dr. Zhou-Run Zhu for the assistance and valuable suggestions provided. This work is supported by National Natural Science Foundation of China under Grants Nos. 12535010, 11935007.

\appendix*
\section{Calibration of the NLO pQCD model calculations}

\begin{figure*}
\begin{center}
\includegraphics[width=0.85\textwidth]{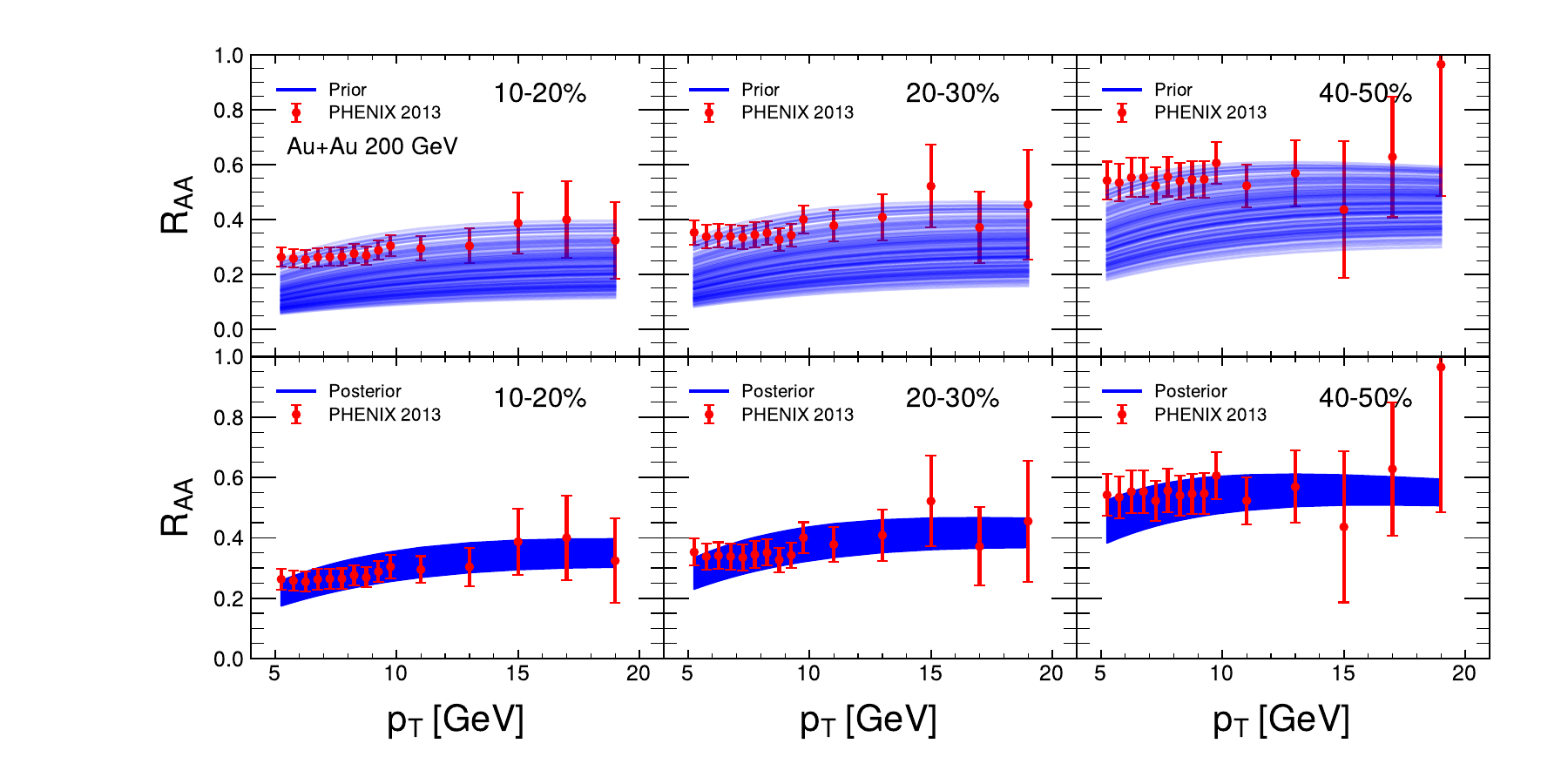}
\includegraphics[width=0.85\textwidth]{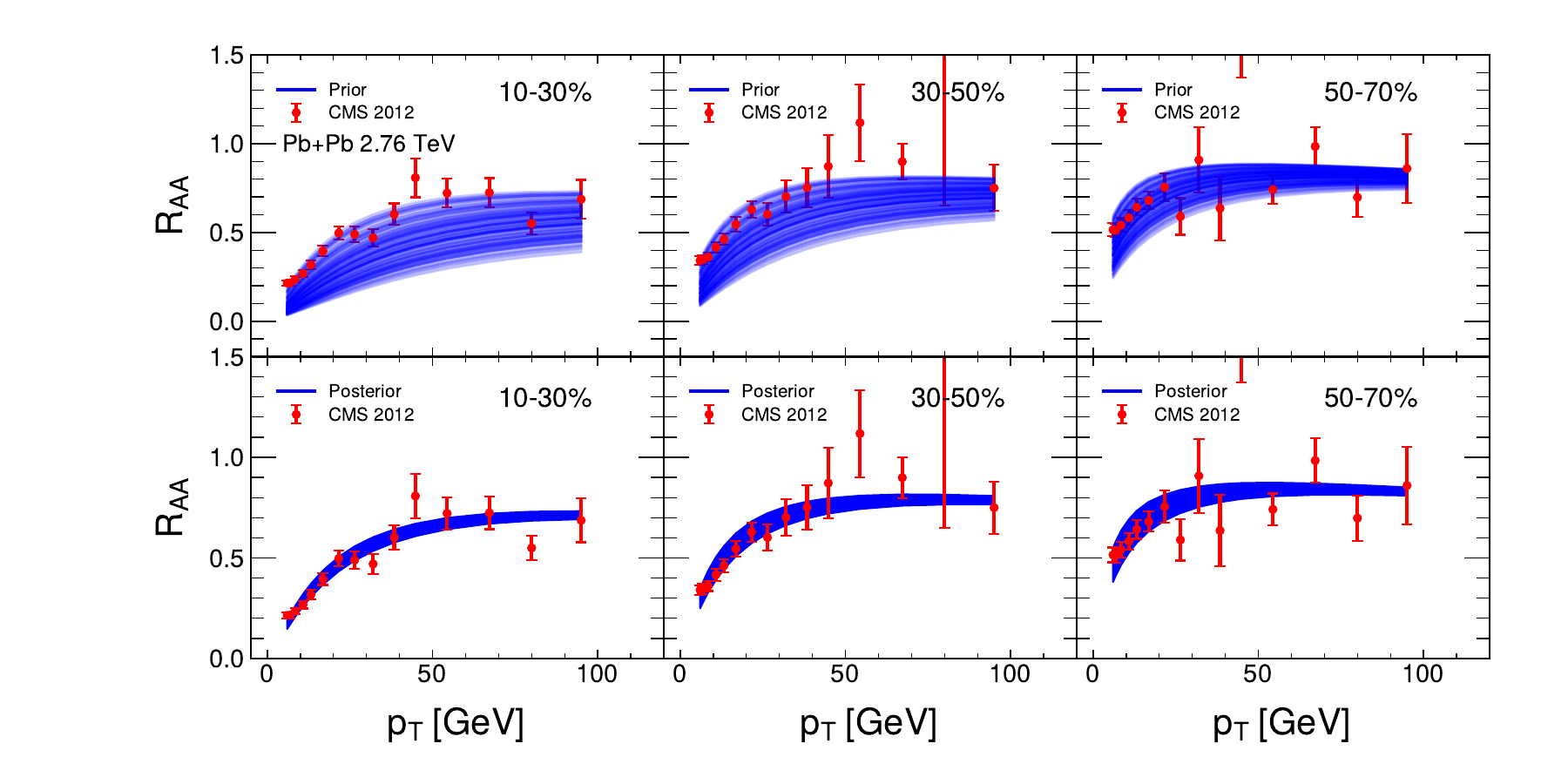}
\includegraphics[width=0.85\textwidth]{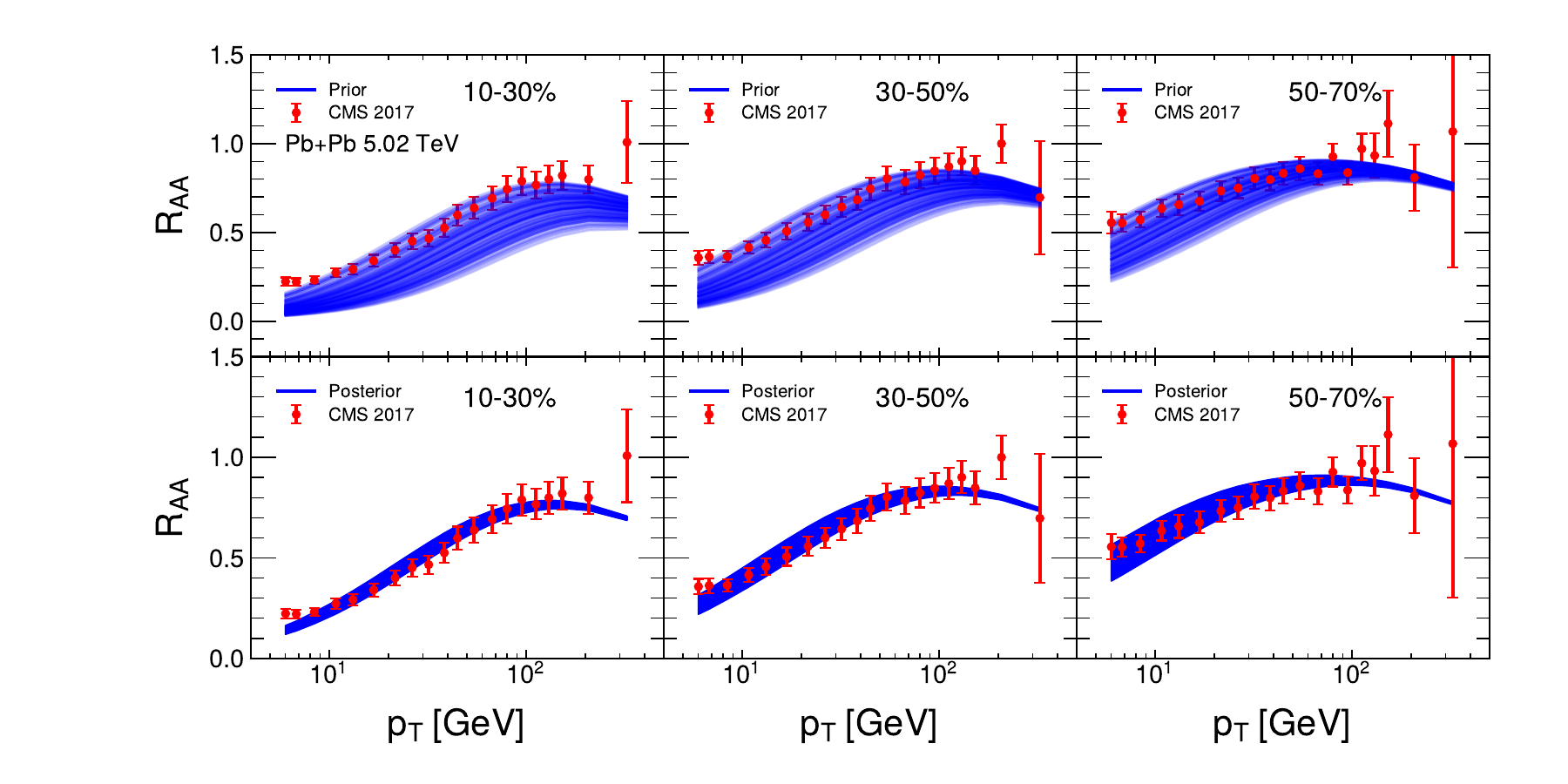}
\end{center}
\caption{Calibration of the NLO pQCD model calculations against the $R_{AA}$ data of charged light hadrons from PHENIX and CMS~\cite{PHENIX:2012jha, CMS:2012aa, CMS:2016xef}. Upper line: 10-20\%, 20-30\%, and 40-50\% Au+Au collisions at $\sqrt{s_{NN}}$ = 200 GeV. Center line: 10-30\%, 30-50\%, and 50-70\% Pb+Pb collisions at $\sqrt{s_{NN}}$ = 2.76 TeV. Lower line: 10-30\%, 30-50\%, and 50-70\% Pb+Pb collisions at $\sqrt{s_{NN}}$ = 5.02 TeV. The upper panel shows calculations using the prior distributions of the magnetic field and chemical potential, while the lower panel displays the posterior distributions after calibration.
}
\label{fig:RAACalibration1}
\end{figure*}

\begin{figure*}
\begin{center}
\includegraphics[width=0.85\textwidth]{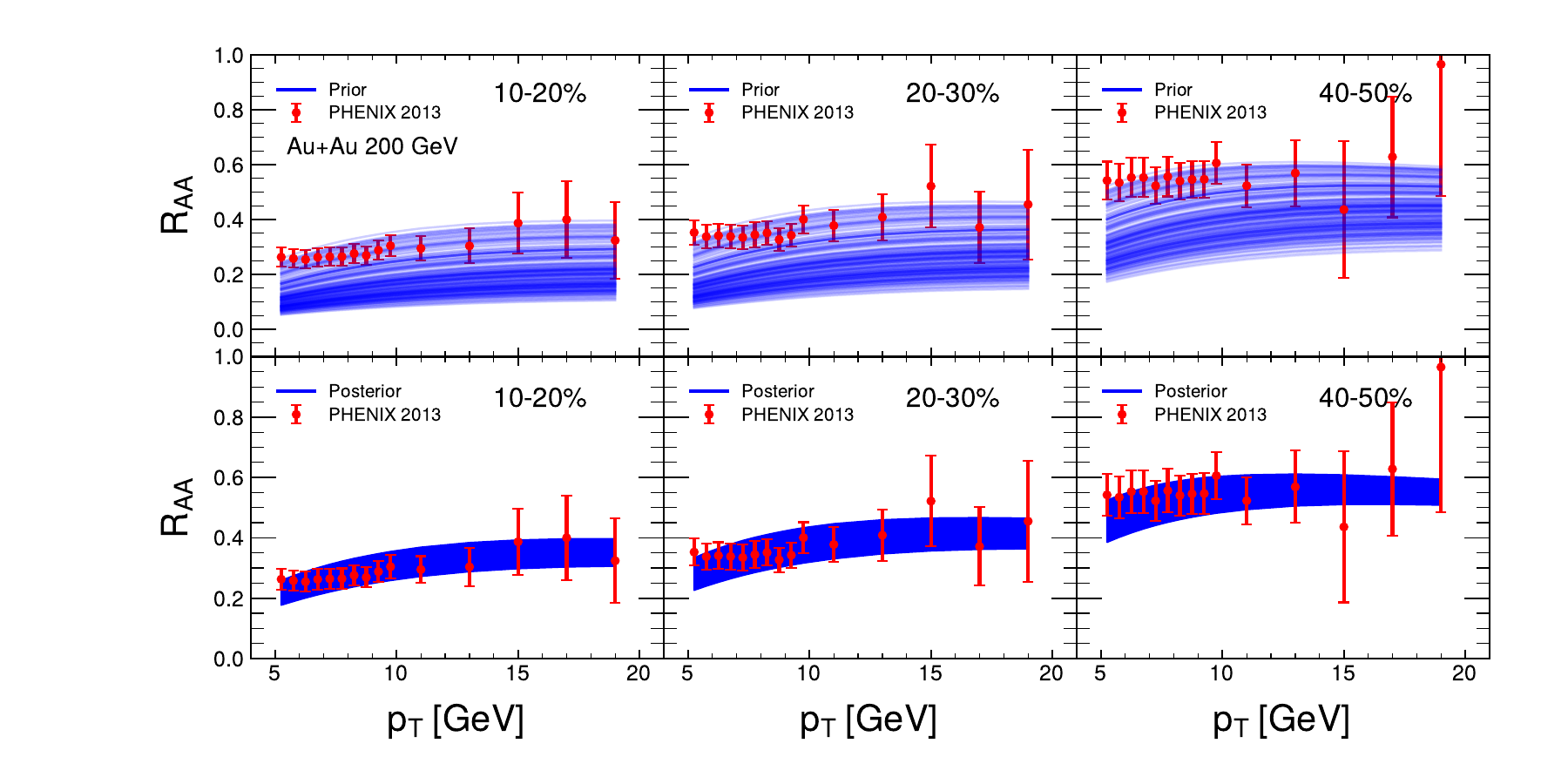}
\includegraphics[width=0.85\textwidth]{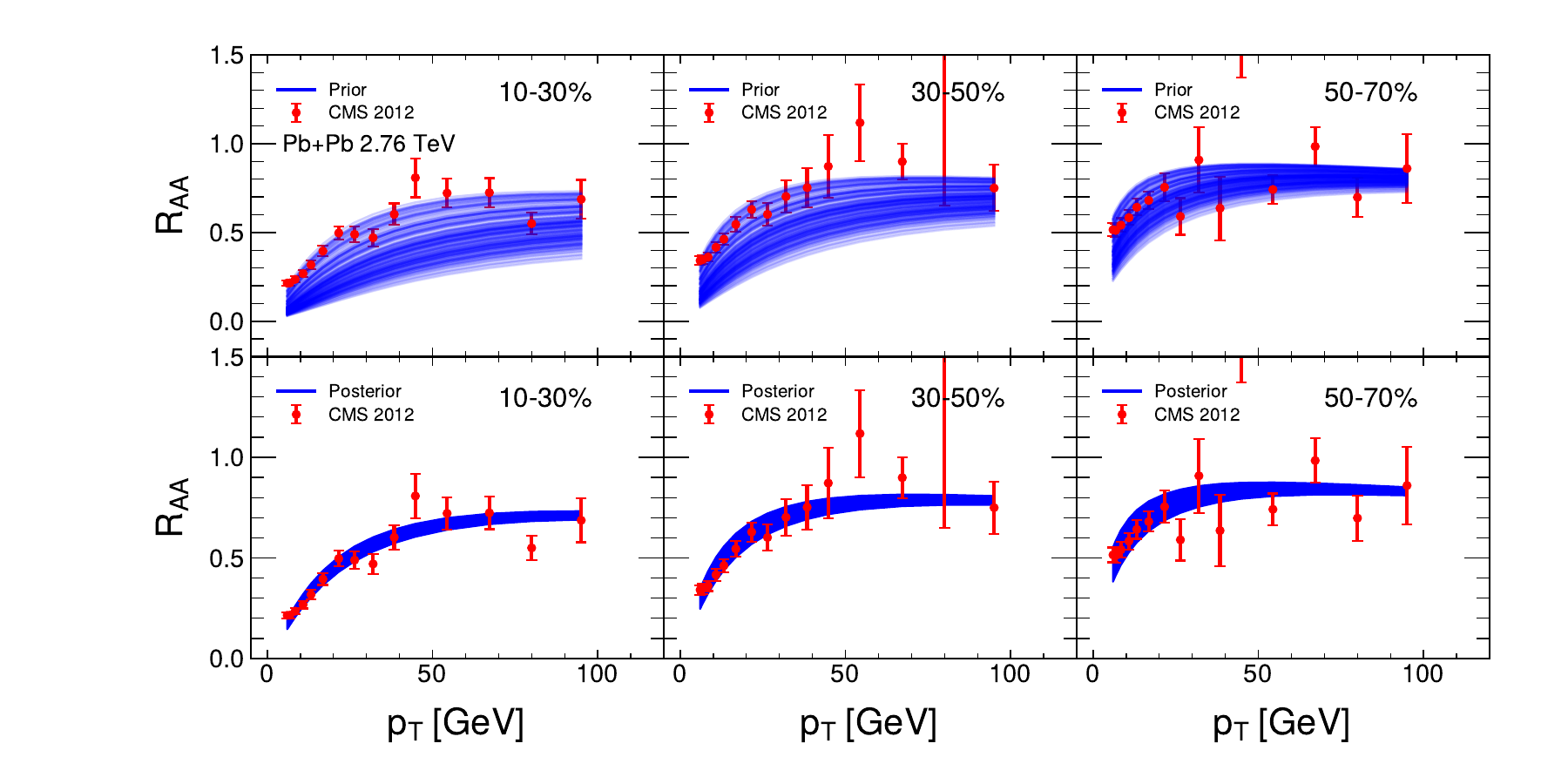}
\includegraphics[width=0.85\textwidth]{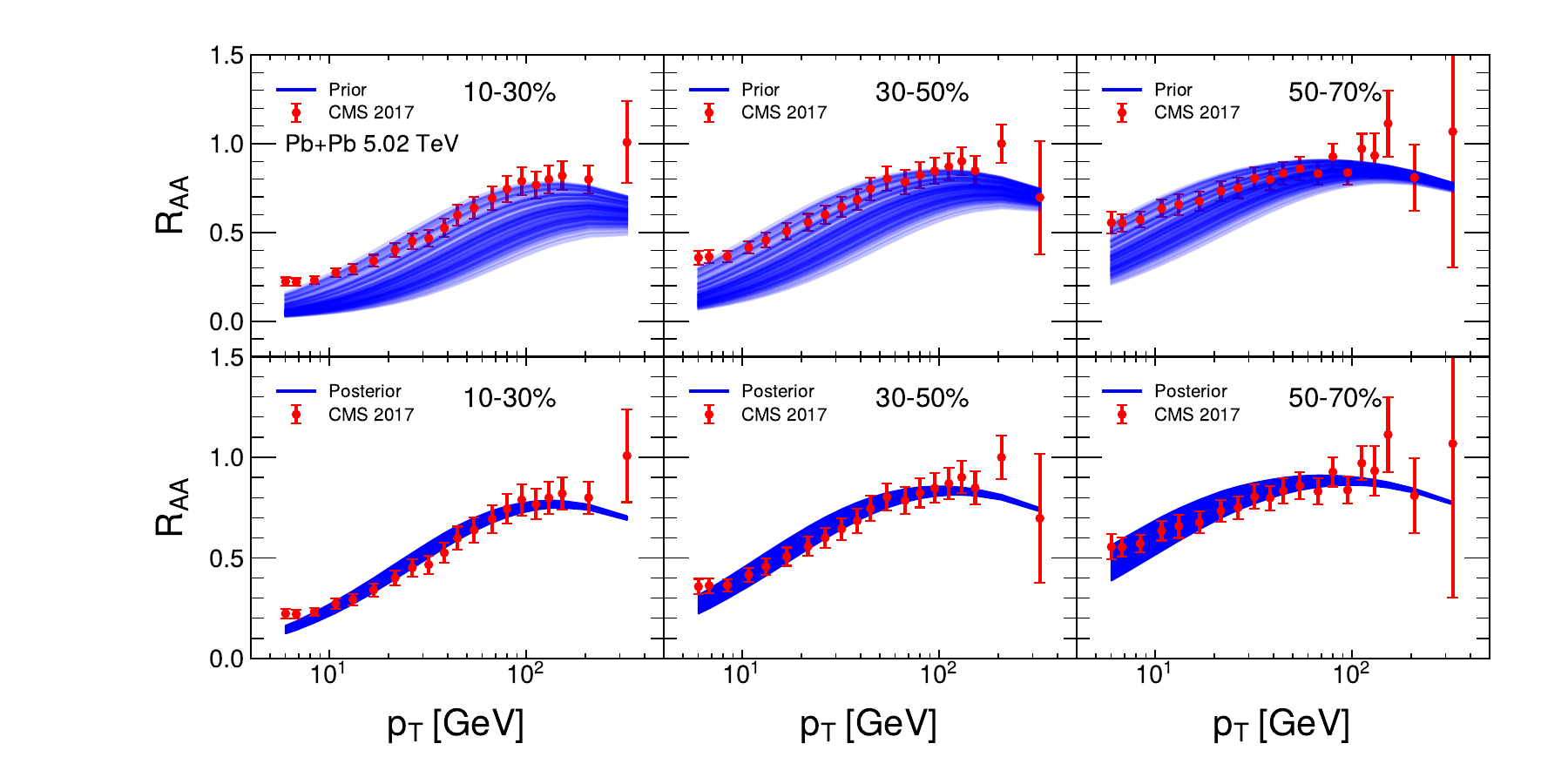}
\end{center}
\caption{Calibration of the NLO pQCD model calculations against the $R_{AA}$ data of charged light hadrons from PHENIX and CMS~\cite{PHENIX:2012jha, CMS:2012aa, CMS:2016xef}. Upper line
: 10-20\%, 20-30\%, and 40-50\% Au+Au collisions at $\sqrt{s_{NN}}$ = 200 GeV. Center line: 10-30\%, 30-50\%, and 50-70\% Pb+Pb collisions at $\sqrt{s_{NN}}$ = 2.76 TeV. Lower line: 10-30\%, 30-50\%, and 50-70\% Pb+Pb collisions at $\sqrt{s_{NN}}$ = 5.02 TeV. The upper panel shows calculations using the prior distributions of the scaled magnetic field and scaled chemical potential, while the lower panel displays the posterior distributions after Bayes inference.}
\label{fig:RAACalibration2}
\end{figure*}

Figure~\ref{fig:RAACalibration1} (based on the first assumption) and figure~\ref{fig:RAACalibration2} (based on the second assumption) respectively present the calculated nuclear modification factor $R_{AA}$
in comparison with experimental data. These two figures illustrate the computational results of the model under different assumptions.
In the figures, the top panels show the prior results, reflecting the model's initial predictive capability without fully utilizing experimental information. The bottom panels display the posterior results obtained through Bayesian inference. It is evident that the posterior results achieve significantly better agreement with the experimental data, demonstrating the optimization of the model after incorporating experimental inputs. This not only validates the feasibility of the modeling method but also further enhances its capability and reliability in describing the actual physical processes. These findings are of great value for gaining deeper insights into the related phenomena.

\bibliographystyle{apsrev4-2}
\bibliography{clv3}

\end{document}